\documentclass[12pt]{iopart}
\usepackage{graphicx}    
\expandafter\let\csname equation*\endcsname\relax
\expandafter\let\csname endequation*\endcsname\relax

\usepackage{amsmath}     
\usepackage{amssymb}     
\usepackage{bm}          
\usepackage{mathrsfs}    
\usepackage{color}       
\usepackage{iopams}      

\usepackage{tabu}
\newcolumntype{"}{@{\hskip\tabcolsep\vrule width 2.5pt\hskip\tabcolsep}}

\newcommand{\tb}{{\textit{T. brucei}}}

\renewcommand{\u}{{\bm u}}

\renewcommand{\r}{{\bm r}}
\newcommand{\J}{{\bm J}}
\newcommand{\vi}{{\bm  v}_i}
\newcommand{\ri}{{\bm r}_i}
\newcommand{\av}[1]{\left\langle #1 \right\rangle}

\begin{document}

\title[\textit{Trypanosoma brucei} moving in microchannels]{\textit{Trypanosoma brucei} moving in microchannels and through constrictions}

\author{Zihan Tan, Julian I. U. Peters, and Holger Stark}

\address{Division of Theoretical Physics, Institute of Physics and Astronomy,
Technische Universit\"at Berlin, Hardenbergstra\ss e 36,10623 Berlin, Germany}
\ead{zihan.tan@tu-berlin.de, holger.stark@tu-berlin.de}

\vspace{10pt}

\begin{abstract}
	\textit{Trypanosoma brucei} (\tb), a single-celled parasite and natural microswimmer, is responsible for fatal sleeping sickness 
in infected mammals, including humans. Understanding how \tb~interacts with fluid environments and navigates through confining spaces
is crucial not only for medical and clinical applications but also for a fundamental understanding of how life organizes in a confined 
microscopic world. Using a hybrid multi-particle collision dynamics (MPCD)--molecular dynamics (MD) approach, we present our 
investigations on the locomotion of an \textit{in silico} \tb\ in three types of fluid environments: bulk fluid, straight cylindrical microchannels, 
and microchannels with constrictions. We observe that the helical swimming trajectory of the \textit{in silico} \tb~becomes rectified in 
straight cylindrical channels compared to bulk fluid. The swimming speed for different channel widths is governed by the diameter of the 
helical trajectory. The speed first slightly increases as the channel narrows and then decreases when the helix diameter is compressed. 
An optimal swimming speed is achieved, when the channel width is approximately twice the bulk helix diameter. It results from an interplay of the trypanosome's hydrodynamic interactions with the cylindrical channel walls and the high deformability of the parasite. In microchannels with constrictions, the motions of the anterior and posterior ends, the end-to-end distance, and the log-rolling motion 
of the cell body are characterized and show salient differences compared to the straight-channel case. Depending on the constriction length 
and width, we observe characteristic slip, stuck, and stuck-slip motions of the model \tb\ within the constriction. Our findings may provide 
some mechanical insights into how \tb\ moves through blood vessels and tissues, and across the blood-brain barrier.
\end{abstract}

%
\vspace{2pc}
\submitto{\NJP}
%
%
%

\section{Introduction}
Swimming microorganisms inhabit complex fluids at low Reynolds number and in confining environments. They often evolve swimming strategies that generate helical trajectories~\cite{Jennings:1901,Purcell:1977} to enhance mobility and environmental adaptability. 
For instance, the helical swimming pattern of spermatozoa (sperm cells) facilitates better sensing of external chemical gradients (chemotaxis)~\cite{Su:2013,Jikeli:2015}, as well as rheotaxis, leading to a stable upstream spiraling motion that aids navigation towards the egg~\cite{Kantsler:2014}. Certain algae, such as \textit{Chlamydomonas}, utilize helical trajectories to optimize light-seeking (phototactic) behavior, which is vital for photosynthesis~\cite{Bennett:2015,Cortese:2021,Leptos:2023}.  Similarly, species of \textit{Tetrahymena}, a genus of free-living ciliates, display helical swimming patterns to orient themselves directly in response to external stimuli~\cite{Marumo:2021}. Finally, also \textit{Escherichia coli} bacteria exhibit helical swimming paths by rotating a bundle of helical flagella and due to the wobbling motion of the cell body \cite{Berg:1973,Silverman:1974,Berg:2004,Hu:2024}.

In the same vein, the single-celled parasite African trypanosome, in particular, the species \textit{Trypanosoma brucei} (\tb)~\cite{Langousis:2014}, 
is another notable example of a microorganism swimming on a helical path. \textit{Trypanosoma brucei} is notoriously responsible for the fatal disease \textit{Trypanosomiasis} (sleeping sickness) in infected mammals. Its helical swimming motion is driven by a beating flagellum helically attached to the spindle-shaped and deformable cell body~\cite{Rodriguez:2009,Koyfman:2011,Heddergott:2012,Alizadehrad:2015}. This allows for robust directional swimming within the bloodstream, even under biological noise~\cite{Wheeler:2017}, and efficient exploitation of complex environments~\cite{Heddergott:2012}. The helical attachment of the flagellum with a half turn around the cell body also optimizes the swimming velocity\ \cite{Alizadehrad:2015}.
The motility of \tb\ is crucial for establishing and maintaining bloodstream infection~\cite{Shimogawa:2018}. However, how the motility adapts to complex environments remains an incomplete picture. A pioneering \textit{in vivo} study in a natural host environment with the species \textit{Trypanosoma carassii} revealed a whip-like motion for changing direction~\cite{Bargul:2016,Doro:2019}, rather than swimming backwards~\cite{Heddergott:2012,Alizadehrad:2015}. In addition, systematic studies in artificial structures that mimic essential features of real environments are performed and thereby add to our current understanding~\cite{Heddergott:2012,DeNiz:2023}. Notably, Sun \textit{et al.} found that \tb\ in flow can squeeze through orifices with a width half of the maximum cell diameter when they move actively~\cite{Sun:2018}. This highlights the role of elastic cell deformations in crossing tight passages.

Understanding how trypanosomes navigate through complex and confining environments is not only of fundamental interest to the physics of microswimmers but also crucial from both clinical and biological perspectives. The deadly sleeping sickness in humans and livestock caused by \tb\ poses severe public health and economic challenges in sub-Saharan Africa~\cite{Simarro:2012}. During its life cycle, the \tb~is transfered by the tsetse fly to the mammalian bloodstream, where it invades the brain, or it is reabsorbed by another tsetse fly. There, it needs to migrate through complex environments adopting different morphotypes~\cite{Alizadehrad:2015,Mogk:2014,Schuster:2017}. Moreover, trypanosomes are not confined to the bloodstream but also inhabit tissues, skin, and organs~\cite{Krueger:2018,Goodwin:1970,Trindade:2016}. Strikingly, evidence suggests that the skin 
may serve as a more preferred habitat than blood vessels~\cite{Capewell:2016}. A recent study on \tb\ infecting 
an artificial human skin revealed that their skin tissue forms could be significant in sustaining infections over extended periods~\cite{Reuter:2023}. 
Additionally, one open question is how \tb\ invades the central nervous system, specifically the brain, by crossing tight junctions such as the blood-brain barrier (BBB)\cite{Mogk:2014,Abbott:2010,Campisi:2018,Park:2019,Hajal:2021}, and/or the blood-cerebrospinal fluid (CSF) barrier\cite{Mogk:2014,Mogk:2017}.

Albeit some studies have been reported, as reviewed above, a comprehensive physical understanding of how \tb\ interacts with complex fluid environments, adapts to, and navigates through confining spaces remains largely elusive. In this context, mesoscopic hydrodynamic modeling can complement 
experimental studies and provide detailed microscopic insights, which are indispensable. To this end, we use an \textit{in silico} model of the African trypanosome, developed in our previous work~\cite{Alizadehrad:2015,Babu:2012}, coupled to a viscous solvent that we 
simulate using the method of multi-particle collision dynamics (MPCD)~\cite{MPCD}. We perform a thorough numerical study for three types of fluid environments using different quantities.
First, in a bulk fluid we demonstrate that the model \tb, with its crescent-like twisted shape, swims on a helical path, for which we determine the diameter. It also performs a rolling motion about its local body axis as observed in experiments\ \cite{Heddergott:2012}. Second, for a set of standard parameters, we find that in channels with decreasing diameter the helical path first aligns with the channel axis. Thereafter, the diameters of the helical paths belonging to the two ends of the cell body sharply drop which indicates that the cell body is straigthened through interactions with the channel walls. Finally and most importantly, we fabricate a constriction within a wide channel. Depending on the length and width of the constriction, we observe three types of trajectories, which are classified in a state diagram: the trypanosome either slips through the constriction or becomes stuck if the channel width approaches the maximum cell-body diameter. For lengths smaller than the cell-body length a stuck-slip motion occurs in between.

This article is structured as follows. Section\ \ref{sec:methods} outlines the mesoscopic hydrodynamic \tb\ model, including the basic framework of the MPCD simulation technique, the required boundary conditions for bulk and channels, and the simulation parameters. Section\ \ref{sec:results} presents the results for swimming in the bulk fluid, through straight channels and constrictions. The resulting helical trajectories are thoroughly analyzed in terms of mean-squared displacement, helical diameters, swimming velocity, rolling motion, end-to-end distance, and retention time. Finally, Sec.~\ref{sec:disc+conc} summarizes the implications of our findings, discusses their relevance, and suggests directions 
for future research.

\section{Simulation models and method}
\label{sec:methods}

In modeling the African trypanosome, we follow the approach outlined in our earlier works in Refs.~\cite{Alizadehrad:2015,Babu:2012}. 
Very recently, this model has also been adapted in fluid simulations using dissipative particle dynamics~\cite{Overberg:2024}. 

In the following we introduce the model \tb\ in depth, explain the basic principles of MPCD to simulate fluid flow initiated by \tb, and then show the geometry of a microchannel with constriction, which is the central focus of our study. We also explain the relevant boundary conditions for our simulations and summarize the parameters.

\subsection{\textit{T. brucei} model} \label{sec:trypmodel}
\begin{figure} 
	\centering
			\includegraphics[width=0.6 \columnwidth,trim={0 0 0 0},clip]{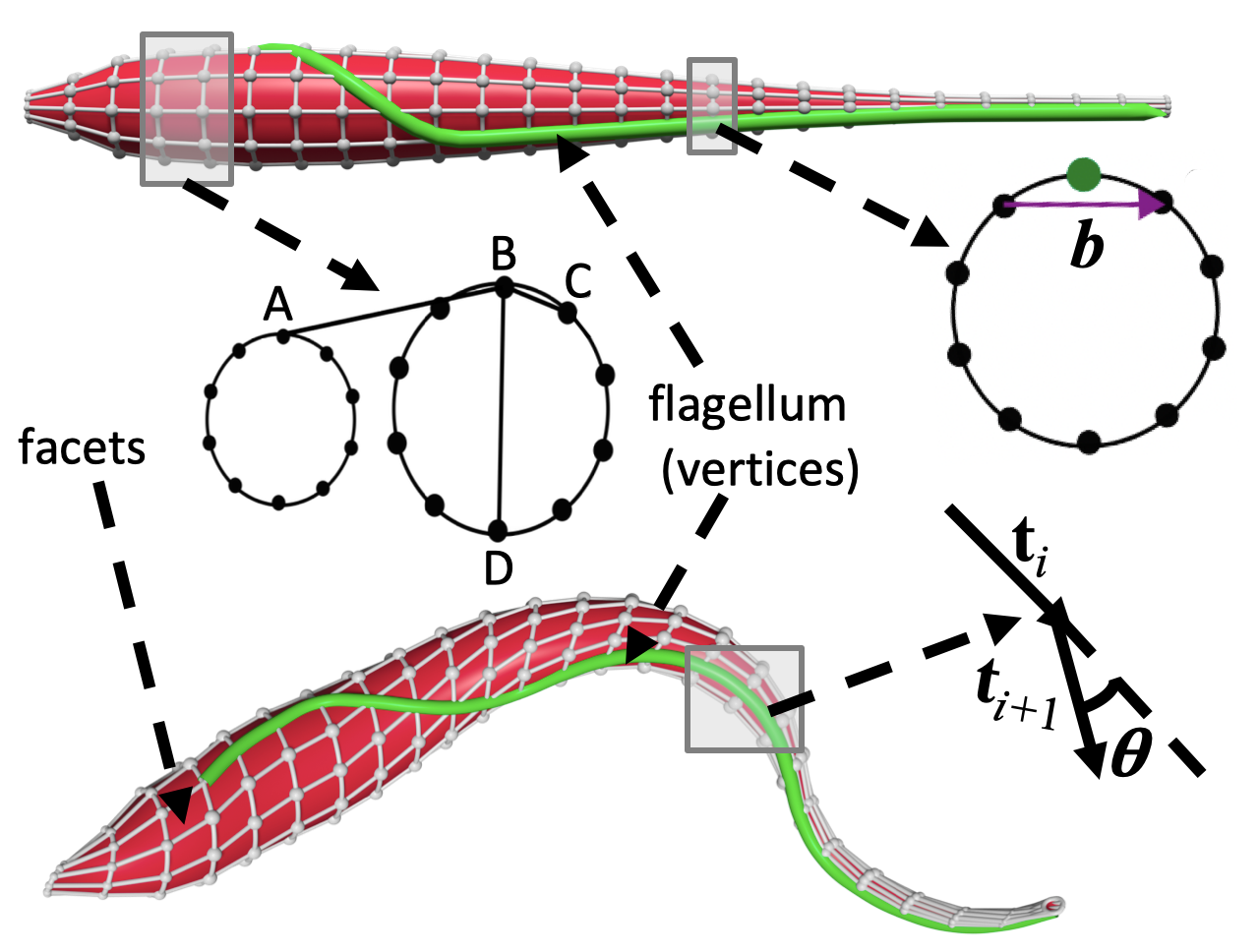}

	\caption{\label{fig:tryp} Schematic representations of the mesoscopic model of a \textit{Trypanosoma} before (top) and during (bottom) deformation due to the bending wave running along the attached flagellum (green).
The straight, spindle-shaped cell body comprises $26$ equally spaced circles arranged along the spindle axis; each circle is 
discretized into $10$ equally spaced
vertices, each with mass $m$. A close-up of two adjacent circles in the middle illustrates the stiff harmonic springs connecting vertices between neighboring circles (A-B) as well as neighboring vertices (B-C) and opposite vertices (B-D) on the same circle. Additionally, bending potentials are applied to each of the ten lines of vertices running from the posterior (thick) to the anterior (thin) ends using tangent vectors $\boldsymbol{t}_i$ along these lines. The flagellum is depicted as a continuous green line, originating from the flagellar pocket (cf. the position of the $4^{\text{th}}$ circle) with a helical half-turn, running along the surface of the cell body and extending to the thinner anterior end. A cross section encloses a flagellum vertex (green disk) and its two nearest vertices defining the bending vector $\boldsymbol{b}$ (see the main text for details).
	}
\end{figure}

Albeit the specific structure of Trypanosoma cells differs between species and life stages, they share many common features at a coarse-grained level. Here, we model the \tb~species in its bloodstream form. Based on observations from experimental microscope images, the geometry of the \tb~body, in the absence of deformation, is modeled with a cylindrically symmetric, spindle-like shape. As sketched in Fig.~\ref{fig:tryp} (top), the cell body, with a total length $L_0=25a_0$ from the thick posterior to the thin anterior end, is divided by 26 segmental rings with equal distance $a_0$ along the long axis of the cell body. Each segment (see close-ups) contains $10$ vertices (represented by small white balls), each with mass $m$, uniformly distributed on the circular segment. Here, $a_0$ and $m$ are the length and mass units related to the MPCD fluid, as described in Sec.~\ref{sec:fluidmodel}. This amounts to a total mass of the \tb~of $M=260m$. The radius of each circular segment, $R(i)$, is determined according to~\cite{Babu:2012}
\begin{align}
	R(i)=&\exp\{-0.006(i-0.85)^2\}\nonumber\\
	&(i-0.85)^{0.25}a_0, \qquad 1\leq i \leq 26\,,
	\label{eq:R_i}
\end{align}
where $i$ denotes the index of the cell circle, starting from the most left (posterior) end in Fig.~\ref{fig:tryp}. Note the formula for $R(i)$ is a phenomenological fit to the real body shape.

To maintain the shape of each segment and the cell body as a whole, the nearest vertices on the circular segment (cf. the representative B-C pair in Fig.~\ref{fig:tryp}) and along the long axis of the undeformed cell body (cf. the representative A-B pair) are connected by a harmonic potential
\begin{align}
	U_s & =\frac{1}{2}k_s(l-l_s)^2 \, .
	\label{eq:usprings}
\end{align}
Here, $l$ is the distance between the nearest vertices and $l_s$ is the corresponding equilibrium value, which can be calculated for each vertex pair from Eq.\ (\ref{eq:R_i}) and the distance of the segments.
Furthermore, to keep the circular segments stable, the opposite vertices on the segment (cf. the representative B-D pair) are also connected by the spring potential of Eq.~\eqref{eq:usprings}.
The spring constant $k_s$ is set to very high values for all springs (see Table \ref{tb:parameters}) to keep the distances $l$ close to their equilibrium values $l_s$.

Furthermore, the cell body of the \tb\ possesses bending stiffness through a cortex of microtubules running from the posterior to the anterior 
end along the long axis of the cell body \cite{Heddergott:2012,Babu:2012,Hemphill:1991}.
%
%
To model this resistance to bending deformations, the following bending potential is applied to the $10$ lines running along the long axis of the cell body, 
\begin{align}
	U_b&=\frac{1}{2}\kappa_b(\cos\theta-\cos\theta_0) ,
	\label{eq:bend}
\end{align}
where $\kappa_b$ is the bending rigidity, and $\theta$ and $\theta_0$ are the actual and equilibrium angles between two adjacent bond vectors $\boldsymbol{t}_i$ and $\boldsymbol{t}_{i+1}$, respectively. Here, $\boldsymbol{t}_i=\r_{v,i+1}-\r_{v,i}$, where $\r_{v,i}$ denotes the position of the vertex on a described line and the $i^{\text{th}}$ circle. Thus, $\cos\theta=\boldsymbol{t}_i\cdot\boldsymbol{t}_{i+1}/|\boldsymbol{t}_i||\boldsymbol{t}_{i+1}|$. The equilibrium angle of pre-bending, $\theta_0$, is calculated from $R(i)$ and the distance of the segments. We assign a local bending rigidity depending on the cross-section area of the cell body for the first $21$ relevant
segments counting from the posterior end~\cite{Alizadehrad:2015}. Specifically, $\kappa_b=\kappa_{b0}A_i/A_0$, 
where $2 \le i \le 21$, $\kappa_{b0}$ is the mean bending rigidity, and $A_0$ the mean cross-sectional area. For the last $4$ involved segments towards the anterior end, we let $\kappa_{b}$ follow a power-law reduction, $\kappa_{b} = \kappa_{b0} \, 0.8^{i-22}$ ($22 \le i \le 25$),
making the anterior part of the cell body more flexible \cite{Alizadehrad:2015}. This is motivated by the experimental 
observations \cite{Heddergott:2012,Alizadehrad:2015} that the anterior tip typically makes larger excursions than the other parts
of the cell body. Furthermore, not all microtubules run to the very end of the anterior tip and the flagellum itself runs a bit beyond the
cell body. So the reduction in the bending rigidity is well justified.
However, $\kappa_b \propto A_i$ is a pure phenomenological choice to quantify that thicker parts of the cell body bend less due to a larger
bending rigidity.

Moreover, the \tb\ body is effectively described as a closed, impermeable (i.e., fluid cannot penetrate it), but deformable shell (see red surfaces in Fig.~\ref{fig:tryp}), with two hemispheres at each pole (not drawn). The high value of the spring constant $k_s$ ensures that the cell body effectively satisfies both surface and volume constraints. In fact, we checked that both volume and surface area of the modeled \tb\ varies by less than $3\%$ during its motion in the bulk fluid and while passing through constrictions.

A flagellum, prescribed by $23$ vertices (out of the total 
$260$ vertices) on the cell body and connected by bonds, is depicted as the solid green line in Fig.~\ref{fig:tryp}. It originates at the 
flagellar pocket (a selected vertex on the $4^{\text{th}}$ circle at the posterior left side) and runs longitudinally to the $7^{\text{th}}$
circle. From there, it wraps along the diagonals of the net elements around the body in a half-circle trajectory (enclosing $6$ vertices, shown as the solid green wire) and then extends straight again from the $12^{\text{th}}$ circle to the thinner anterior end at 
the $26^{\text{th}}$ circle. 

The activity of the flagellum is simulated by a sinusoidal bending-wave potential propagating along 
its length. This active potential deforms the entire cell body to a ``\textit{crescent}"-like chiral shape and, through interactions with the surrounding fluid, breaks temporal-spatial symmetry to generate propulsion. Explicitly, the bending wave potential imposed on the flagellum is given by
\begin{align}
	U_w &= \frac{1}{2} \kappa_w \left[ \boldsymbol{t}_{j+1}-\boldsymbol{\Re}(l_j\beta_w)\boldsymbol{t}_{j}\right]^2,
	\label{eq:Uw}
\end{align}
where $\kappa_w$ is the bending rigidity of the flagellum, $\boldsymbol{t}_j$ denotes the bond vector connecting two adjacent flagellum vertices and pointing towards the anterior end, and the subscript $j$ represents the index of the flagellum vertex. Taking the flagellum attached to the undeformed straight cell body (cf. the top panel in Fig.\ref{fig:tryp}) as the reference state, the bending potential in Eq. \eqref{eq:Uw} is applied by rotating $\boldsymbol{t}_j$ about the local surface normal $\boldsymbol{n}$
by an angle $l_j\beta_w$ with $l_j=|\boldsymbol{t}_j|$ and $\boldsymbol{\Re}$ is the corresponding rotation matrix. The rotation angle $\beta_w$ varies in a sinusoidal fashion with amplitude $\beta_0$, explicitly~\cite{Alizadehrad:2015,Babu:2012,Yang:2008} 
\begin{align}
	\beta_w &= \beta_{0} \sin \left[ k d_j+ \omega t\right],
	\label{eq:betaw}
\end{align}
where $d_j$ measures the distance from the flagellum pocket vertex (at the $4^{\text{th}}$ circle) to its vertex $j$. The wave 
number $k=2\pi/\lambda$ contains the wave length $\lambda$, which is $0.428L_0$ in our simulations, and $\omega$ is the angular frequency.
Finally, we note that the local body or surface normal is defined as 
$\boldsymbol{n}=\boldsymbol{t}_j \times \boldsymbol{b} / |\boldsymbol{t}_j| | \boldsymbol{b}|$, where $\boldsymbol{b}$ 
is a bending vector determined by two adjacent vertices in the ring segment around the $j^{\text{th}}$ flagellum vertex 
(see Fig.\ \ref{fig:tryp}). To further enhance the planar beating pattern, starting from the $13^{\text{th}}$ flagellum vertex, we copy the 
bending vector $\boldsymbol{b}$ to all following vertices towards the anterior end.

\subsection{Fluid model: Multi-particle collision dynamics}\label{sec:fluidmodel}
To account for the fluid flow initiated by the swimming \tb~that is crucial for parasite swimming and for hydrodynamic interactions with complex bounding surfaces, we simulate the fluid environment using multi-particle collision dynamics (MPCD). MPCD solves the Navier-Stokes equations at a particle-based level and
intrinsically captures thermal fluctuations~\cite{MPCD,kap99,kap00,kapral_review,Ripoll:2005}. It has proven to be an efficient and versatile method in soft matter and biophysical research, with successful applications spanning from
colloids~\cite{Padding:2004,Ripoll:2008,Franosch:2011,tan01,tan03}, 
proteins~\cite{Echeverria:2012,Bucciarelli:2016}, 
blood cells and vesicles~\cite{McWhirter:2009,
Dasanna:2019,Noguchi:2005}, 
passive and active polymers~\cite{Weiss:2019,Huang:2010,Lamura:2019,Chelakkot:2012,ClopesLlahi:2022,Jaiswal:2024},  
microfluidic systems~\cite{Yang:2016,tan02}, and
microswimmers~\cite{zottl2014,Hu:2015,Blaschke:2016,Rode:2019,Rode:2021,Zantop:2022,Roca:2022,Clopes:2022,Ruehle:2020,Ning:2023,McGovern:2024}, 
to name a few. 
The MPCD fluid consists of
$N$ point particles each with mass $m$, typically enclosed in a simulation box of dimensions $(L_x,L_y,L_z)$ in three Cartesian coordinates, subject to periodic boundary conditions (PBCs). 
The dynamics of the fluid particles evolves through discrete streaming and collision steps. In the streaming step, the particles move 
ballistically over a time interval $h$, known as the collision time, which effectively sets the mean free time. Accordingly, the position 
$\ri$ of a fluid particle $i$, with $i\in\{1,\ldots,N\}$, is updated according to 
\begin{align} 
	\ri(t+h)=&\ri(t)+h \u_i(t)\,,
	\label{eq:stream}
\end{align}
where $\u_i$ is the particle velocity. 

In the subsequent collision step, the MPCD particles are grouped into cubic cells of edge length $a_0$ defining the local interaction environment (collision cells) and their momenta are redistributed such that the total momentum in the collision cell is conserved. We employ the MPCD variant with the Anderson thermostat (AT) and angular momentum conservation, referred to as  MPCD-AT+a~\cite{zottl2014,Blaschke:2016,Allahyarov:2002}. In this scheme, the particle velocity $\u_i$ updates according to~\cite{MPCD,Zantop:2022}
\begin{align}
	\u_i(t+h)=& \u_\text{cm}(t)+ \delta{\u}_i^\text{ran}	-\Delta\u_\text{cm}\nonumber \\
	&-\tilde{\r}_i (t)\times \boldsymbol{I}^{-1}_\text{cm}(\J_\text{cm}-\Delta \J_\text{cm})\,.
	\label{eq:collide}
\end{align}
Here, $\u_\text{cm}$ is the center-of-mass velocity and
$\delta{\u}_i^\text{ran}$ is a random relative velocity drawn from a three-dimensional (3D) Maxwell-Boltzmann distribution of variance $\sqrt{k_BT/m}$ with $k_BT$ the thermal energy and $k_B$ the Boltzmann constant. The third term $\Delta \u_\text{cm}$ on the right-hand side of Eq.~\eqref{eq:collide} restores linear-momentum conservation. In the last term, $\boldsymbol{I}_\text{cm}$ is the moment-of-inertia tensor of all the particles within the same cell calculated relative to the center of mass and $\tilde{\r}_i=\ri-\r_\text{cm}$ is the relative position of particle $i$ in the particular collision cell with respect to the center-of-mass position $\r_\text{cm}$ of the cell. Inside the bracket, the difference between the angular momentum before the collision, $\J_\text{cm}$, and the change in angular momentum, $\Delta \J_\text{cm}$, is computed to guarantee angular-momentum conservation. Explicitly,
\begin{align}
	\Delta\u_{\text{cm}}& = \sum\limits_{i \in N_{\text{cell}}} \delta \u_i/N_{\text{cell}},\\
	\J_{cm} & =\frac{m}{N_{cell}}\sum_{i \in N_{\text{cell}}} (\tilde{\r}_i\times\u_i) ,\\
	\Delta\J_{cm}&=\frac{m}{N_{cell}}\sum_{i \in N_{\text{cell}}} (\tilde{\r}_i\times\delta\u_i).
	\label{eq:ang_mo}
\end{align}
Here, $N_{\text{cell}}$ denotes the number of particles inside the collision cell. 

The discretization into collision cells breaks Galilean invariance, which is restored by randomly shifting the collision cell lattice at every collision step \cite{ihl03}. Effectively, MPCD-AT serves a dual role as both a collision procedure and a thermostat, enabling simulations in the canonical ensemble without requiring additional velocity rescaling, even in non-equilibrium scenarios.

Moreover, the coupling between fluid and \tb~motions is realized by integrating the \tb~vertices into the MPCD collision step (Eq.~\eqref{eq:collide}). The dynamics of the vertices of the \tb~cell due to the forces acting on them via the spring and bending potentials are integrated using a classical velocity Verlet scheme~\cite{allen}, but with a time step $\Delta t$ much smaller than $h$, typically $h = 400\Delta t$.
\subsection{Microchannels and boundary conditions}\label{sec:bcmodel}
\begin{figure}[t]
	\centering
			\includegraphics[width=0.8\columnwidth]{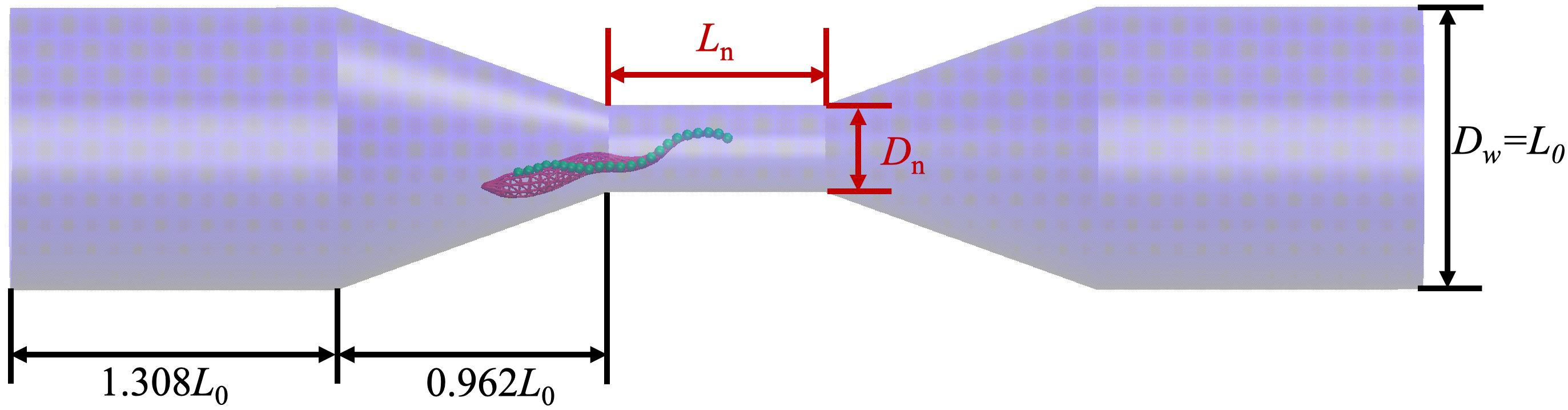}
	\caption{\label{fig:cone} A schematic depiction of an \textit{in silico} \textit{T. brucei} in a microschannel with a constriction. It consists of two pieces of a wide cylindrical section (with diameter $D_w=L_{0}$), two cone-shaped sections, and a narrow cylindrical section as constriction with variable length $L_n$ and diameter $D_n$. The lengths of the wide and cone-shaped sections are fixed. For $D_n=D_w$ a straight cylindrical channel is recovered.}
\end{figure}
\subsubsection{Microchannels with constrictions}
To investigate the dynamics of a \tb~in a microchannel with a constriction region, we design a cylindrical microchannel featuring cone-shape transition zones, as shown in Fig.~\ref{fig:cone}~\cite{Chelakkot:2012}. We choose periodic boundary conditions along the channel axis in $x$ direction. The constriction is connected to the wider channel parts, which have a fixed length $L_w = 1.308L_0$ and width $D_w = L_0$, using a cone-shaped transition zone with fixed length $L_{\text{cone}}=0.962L_0$. In our simulations, we vary the length and width of the constriction channel, denoted as $L_n$ and $D_n$, respectively.
Note, for $D_n = D_w$ the straight-cylindrical-channel configuration is recovered.

\subsubsection{No-slip hydrodynamic boundary conditions}
The so-called random reflection rule between MPCD solvent particles and bounding surfaces is implemented to satisfy no-slip boundary conditions~\cite{Alizadehrad:2015,Babu:2012,padding2005stick,padding06,OlartePlata:2019}. The central idea is that the MPCD solvent particles proceed with their dynamics for a full MPCD time step $h$. During this step, we detect and calculate the intersection time $\delta \cdot h$ ($\delta \in (0,1)$), when a particle collides with the surface at $t + \delta \cdot h$. These particles are then reflected by assigning them a random velocity relative to the velocity of the wall or the nearest cell-body vertex. The tangential component of this velocity is sampled from a Maxwell-Boltzmann distribution, while the normal component is drawn from a Rayleigh distribution. Alternatively, no-slip hydrodynamic boundary conditions in MPCD can also be realized using a bounce-back rule in combination with virtual particles~\cite{tan03,tan02,Rode:2019,padding2005stick,padding06,Muench:2016,theers2016modeling,Kai:2022}.

In our simulations, the exact intersection time is computed and applied for particle reflections in straight cylindrical channels. For geometrically complex boundaries, in our case the channel with the constriction and cone-shaped regions, the above no-slip boundary is simplified. Instead of calculating the exact intersection point between the fluid particle trajectory and the wall, the velocity of a particle $\u_i$ hitting the wall during time $h$ is reflected near the wall directly at time $t$. Specifically, the random reflection is applied within a layer of thickness $\u_i \cdot \boldsymbol{n} h$, where $\u_i \cdot \boldsymbol{n}$ represents the normal component of the particle velocity relative to the surface. The error introduced by this approximation is minimal. As fluid velocities are of the order of one, the layer thickness becomes much smaller than $a_0$~\cite{Rode:2019}.

When accounting for particles colliding with the \tb~cell, we further speed up the simulations by using a common value $\delta \cdot h=h/2$ for all the colliding MPCD particles, rather than calculating the individual collision events at the exact times $t + \delta \cdot h$ of each fluid particle-\tb~collision. This simplification maintains nearly the same level of accuracy as when exact $\delta \cdot h$ values are used, especially for small collision time steps~\cite{Alizadehrad:2015,Babu:2012,padding2005stick,theers2016modeling,hecht2005}. Finally, the total linear momentum exchanged during these collisions is summed up and evenly distributed among the \tb~vertices. Given that the \tb~interacts with numerous MPCD solvent particles during each time interval $h$, this statistical average approximation is shown to excellently capture the essential features of the swimming \tb~\cite{Babu:2012}.

\subsubsection{Bounce-forward boundary conditions: \tb~vertex-wall collision}
When a \tb~vertex of index $i$ is detected to overlap with the channel wall, the so-called bounce-forward approach is employed to handle the collision events such that the \tb~can slide along the wall~\cite{zottl2014,Muench:2016,Zoettl:2012,Schaar:2015}. Compared to the solvent particles, a much smaller MD time step $\Delta t = h/400$ is used for the \tb~vertices, allowing us to approximate the intersection time accurately.

First, the \tb~vertex is moved back in time by half of the MD time step, $0.5\Delta t$, to retreat from the wall. Next, the vertex position is moved onto the surface along the surface normal. The normal component of the vertex velocity is then reversed, resulting in a post-collision velocity given by $\vi' = \vi - 2(\vi \cdot \boldsymbol{n})\boldsymbol{n}$, where $\boldsymbol{n}$ is the local unit normal vector at the wall. Finally, the vertex advances forward using the updated velocity $\vi'$ for the remaining half of the MD time step.

\subsection{Parameters and physical units}\label{sec:parameters}
Here and in what follows, for simulating the MPCD fluid, masses, lengths, and energies are measured in units of our simulation parameters $m$, $a_0$, and the thermal energy $k_BT$. The average number of particles per collision cell is selected as $\av{N_c}=10$ and the collision time step is set to $h=0.01$ in MPCD time unit $t_0 = \sqrt{m a_0^2 / (k_BT)}$. With a mass density $\rho = \langle N_c \rangle m / a_0^3$, the corresponding kinematic viscosity is $\nu = 3.59\nu_0$~\cite{MPCD}, where $\nu_0 = \eta_0 a_0^3 / m$ and $\eta_0 = \sqrt{m k_BT} / a_0^2$. The related dimensionless Schmidt number is $\text{Sc} =\nu/d_0=478.6$ and it  indicates that the viscous diffusion of (transversal) momentum in the fluid is distinctly faster than diffusive mass transport, with the latter characterized by the mass diffusion coefficient $d_0$ of the fluid particles~\cite{MPCD}.

\begin{table}
	\caption{Typical values of the mechanical parameters for the \tb~model. Note that $t_0=\sqrt{ma_0^2/(k_BT)}$ is the time unit 
		used in the MPCD fluid.
	}
	\centering
	\renewcommand{\arraystretch}{1.5}
	\begin{tabular}{|c|| c | c | c | c|c|c|}
		\hline 
		\bf{param.}  & $\beta_0a_0$ &$\omega t_0$&$k_s a_0^2/k_BT$&$\kappa_{b0}/k_BT$&$\kappa_wa_0^2/k_BT $ \\
		\hline 
		\bf{values} &$1.0$ & $0.038$	& $3.0\times10^7$ & 	$3.5\times10^4$ &	$5\times10^3$\\
		\hline
	\end{tabular}
	\label{tb:parameters}
\end{table}

Simulations in the bulk fuid are performed using periodic boundary conditions to a cuboid simulation box of dimensions $(L_x,L_y,L_z)/a_0 = (100,60,60)$. For simulations in straight cylindrical channels, $L_x/a_0 = 100$.
Unless specified, we use the mechanical parameter values listed in Table~\ref{tb:parameters} for the \tb~cell, which gives a shape close to the realistic ``\textit{crescent}"-like form during swimming. To interpret our simulation results, the length, time, and velocity are expressed in units of $L_0$ or $D_0 = 2.574a_0$ (with $D_0$ representing the maximum diameter of the $26$ circles), $t_\omega = 2\pi / \omega$, and $v_\omega = \omega L_0$, respectively.

Finally, we map the simulation units of the MPCD method and the model trypanosome to physical units to give an idea, what our parameters mean. Following Refs.~[18,38], we take the typical length of a \textit{T.\ brucei} as $25\mu\text{m}$, which corresponds to the length of our in-silico model trypanosome of $25 a_0$.
	This yields a length unit of $a_0 = 1.0 \mu\text{m}$.
	We assume the density of the MPCD fluid matches that of water, namely $\rho_{\text{water}} = \langle N_c \rangle m / a_0^3 = 1000\text{kg/m}^3$, resulting in a mass unit of $m = 10^{-16}\text{kg}$. Considering room temperature $T \sim 300\text{K}$ for the studied trypanosomes, the MPCD time unit in physical units is estimated as $t_0=a_0\sqrt{m/(k_BT)}= 1.55 \times 10^{-4}\text{s}$. The corresponding kinematic viscosity of the MPCD fluid is $\nu = 3.59\nu_0 \approx 2.31 \times 10^{-8}\text{m}^2/\text{s}$, which is approximately $43$ times smaller than that of real water. Similar parameter choices have been made
in Refs.~\cite{Babu:2012,Yang:2016,tan02,Hu:2015}.

\section{Results}
\label{sec:results}

We present our simulation results for the \textit{in silico} \tb\ swimming in bulk, through a straight cylindrical channel, and a channel with constriction.

\subsection{Swimming in bulk fluid}
\label{subsec:bulk}

\begin{figure}
			\hspace{2mm}
	\raggedright
				\includegraphics[width=0.45\columnwidth,trim={0 0 0 0},clip]{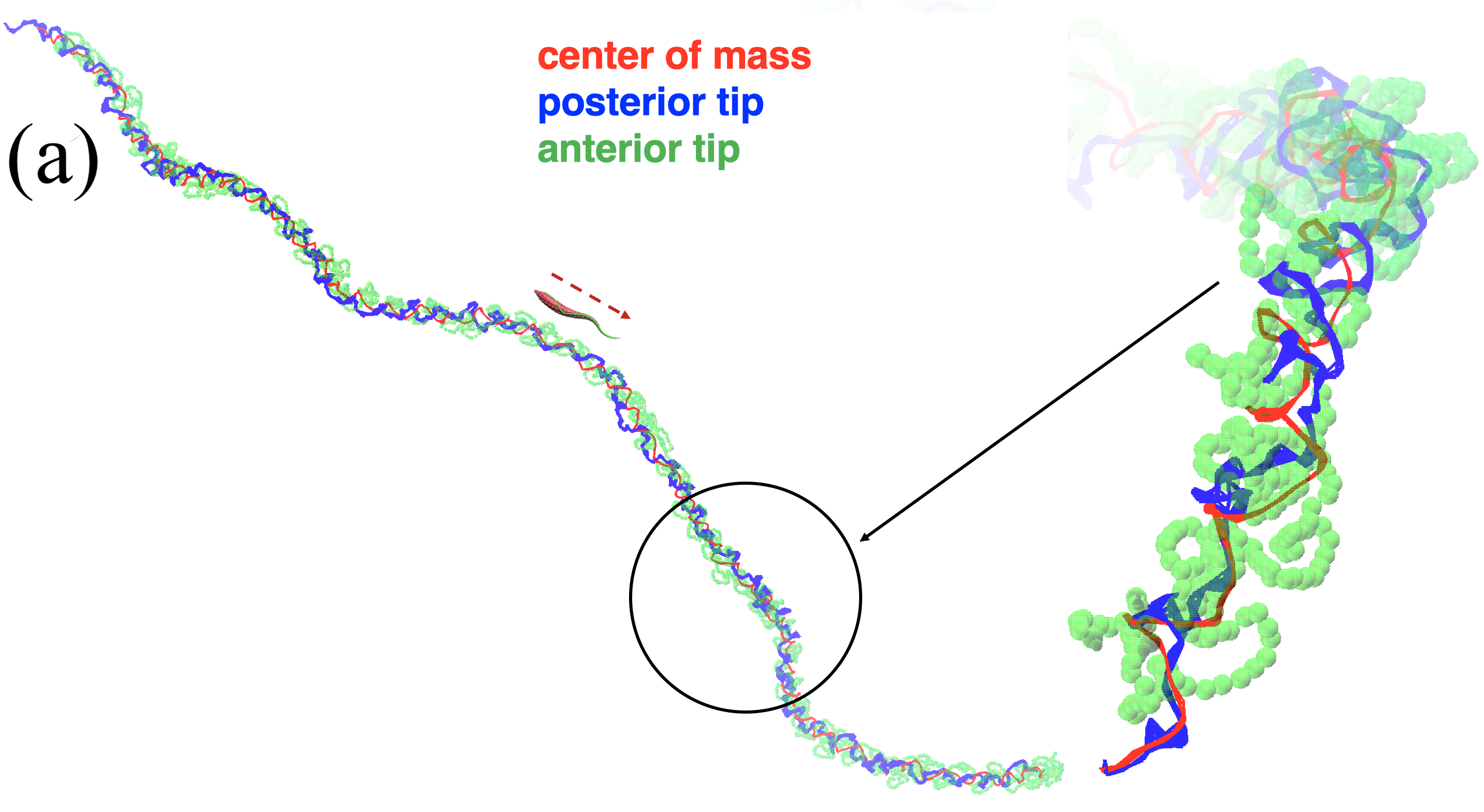}\\
		\vspace{-36mm}
		\raggedright
	\includegraphics[width=0.475\columnwidth,trim={0 0 0 0},clip]{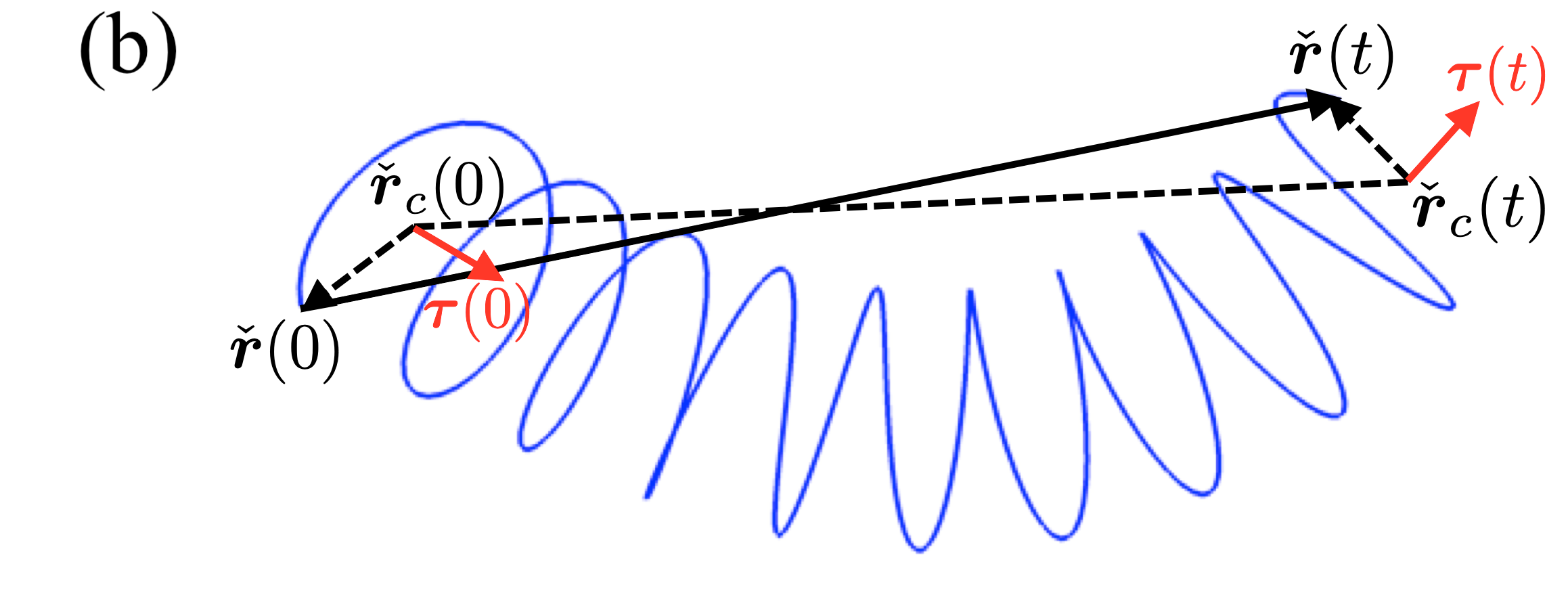}
		\hspace{1.1mm}
					\centering \vspace{-2mm}
	\includegraphics[width=0.5\columnwidth,trim={0 0 0 0},clip]{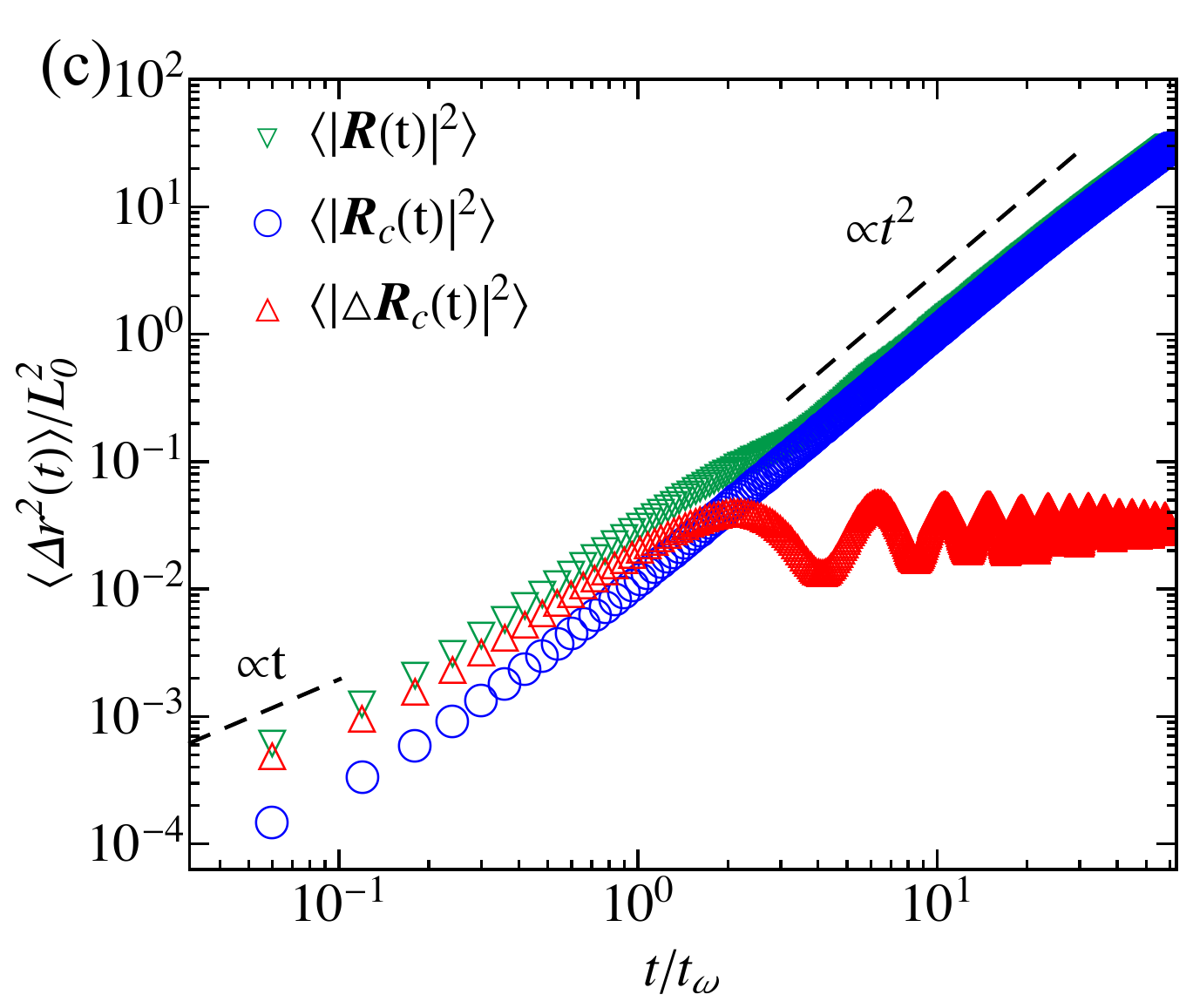}		
	\caption{\label{fig:bulk}(a)~A representative helical swimming trajectory of a swimming \textit{T. brucei} obtained from a MPCD simulation run is shown, with an amplified segment. The center of mass (CM), posterior and anterior are colored in red, green, and magenta, with estimated average diameters of $2.02D_0$, $2.41D_0$, and $2.8D_0$ respectively. Periodic boundary conditions applied along the $x$, $y$, and $z$ directions are unwrapped. The \textit{T. brucei} cartoon (with a dashed red arrow indicating its swimming direction) serves as a visual guide, highlighting the overall length of the nearly straight trajectory. (b) Helical trajectory with the position vector $\check{\boldsymbol{\r}}(t)$, the helix centerline described by the vector $\check{\boldsymbol{\r}}_c(t)$ and the local tangent vector $\boldsymbol{\tau}(t)$ of the centerline. (c)~Time evolution of the MSDs for the CM $\langle|\boldsymbol{R}(t) |^2\rangle$ (green triangles), the helix centerline $\langle|\boldsymbol{R}_c(t) |^2\rangle$ (blue circles), and the relative motion $\langle| \Delta \boldsymbol{R}_c(t) |^2\rangle$ (red triangles) are shown. See main text for detailed definitions.

	}
\end{figure}

We first investigate the swimming properties of the \tb\ in an unbounded bulk fluid. Simulations are performed using the model 
parameters described in Table~\ref{tb:parameters} and with periodic boundary conditions (PBCs) applied along the three Cartesian directions.  Due to the partially chiral wrapping of the beating flagellum around the cell 
surface\ \cite{Rodriguez:2009,Heddergott:2012}, the model \tb\ exhibits a helical swimming trajectory, while the cell body undergoes a ``log-rolling" motion. A similar swimming path has been observed for multi-ciliated microswimmers in Ref.\ \cite{Rode:2021}.
Specifically, Fig.~\ref{fig:bulk}(a)
displays such helical trajectories over a total time $t = 222.6 \, t_\omega$ for the center of mass (CM, red curve), posterior end (blue curve), and anterior tip (green beads) as obtained from one MPCD simulation run. Note that the PBCs are unwrapped in these trajectories to provide a clearer visual impression. A schematic of the model \tb\ together with a dashed red arrow, indicating the swimming direction, is shown alongside to illustrate the scales and that the tra\-jec\-tory length exceeds $20L_0$. 
Figure\ \ref{fig:bulk}(a) reveals two main observations. First, the \textit{in silico} \tb\ predominantly swims in a straight direction, albeit with some curviness. Second, the trajectories of both the CM and the posterior end exhibit clear helical shapes, with the posterior trajectory having a larger diameter, as we will demonstrate below. In contrast, the trajectory of the anterior end demonstrates a more complex path wrapping chirally around the other two curves, as shown in the magnified segment in Fig.~\ref{fig:bulk}(a).
The more complex path is due to the larger excursions of the thinner and more flexible anterior cell end. Quite intriguingly, the centerline of the helical trajectory, \emph{e.g.}, in the upper left corner of Fig.\ \ref{fig:bulk}(a), seems to exhibit a super-helical feature with a much larger pitch and diameter compared to the local helical trajectory, yet maintaining the same chirality.
%
%
We will not investigate this feature further.

Furthermore, we estimate the diameters of the helical trajectories with position vector $\check{\boldsymbol{\r}}(t)$. This needs a careful procedure since the helical path as a whole is not straight but curved. For this, we select a time window $t_\text{helix}$ of (at least) one period of the flagellar bending wave, $t_\omega = 2\pi / \omega$, that slides along the entire trajectory. Within each time window the local center $\check{\boldsymbol{\r}}_c(t)$ and orientation vector $\boldsymbol{\tau}(t)$ of the local helical path
are estimated (see Fig.\ \ref{fig:bulk}(b)). Then, the local diameter of the helix is calculated as the relative distance of the trajectory points to the local centers averaged over the time window, $\langle|\check{\boldsymbol{\r}}(t) - \check{\boldsymbol{\r}}_c(t)|\rangle$. Finally, by averaging over these local diameters, we obtain the effective helix diameter $D_\text{helix}$. Using this method, the computed values of $D_\text{helix}$ for the CM, posterior, and anterior trajectories are $2.02D_0$, $2.41D_0$, and $2.8D_0$, respectively. Below, we will discuss further how these diameters depend on the system parameters.
\subsubsection{Mean-squared displacement}
The helical swimming feature of the \tb\ is also readily unveiled by the mean-squared displacement (MSD) $\langle |\boldsymbol{R}(t) |^2\rangle = \langle |\check{\r}(t) - \check{\r}(0) |^2\rangle$ of the CM. Thus, in the following $\check{\r}(t)$ represents the position vector of the CM. As before, we refer the position vector to the position of the centerline, $\Delta \r(t) = \check{\r}(t) - \check{\r}_c(t)$, and obtain for the displacement in time:
\begin{align}
	\boldsymbol{R}(t) & =  \check{\r}_c(t) - \check{\r}_c(0) + \Delta \r(t) - \Delta \r(0)
	\nonumber \\
	& = \boldsymbol{R}_c(t) + \Delta \boldsymbol{R}_c(t)
\end{align}
This will help us to interpret the motion of the CM. Figure\ \ref{fig:bulk}(c)
presents the MSDs of the CM $\langle|\boldsymbol{R}(t) |^2\rangle$ (shown as green lower triangles), of the helix centerline $\langle|\boldsymbol{R}_c(t) |^2\rangle$ (blue circles), and of the relative motion 
$\langle| \Delta \boldsymbol{R}_c(t) |^2\rangle$ (red triangles).
The MSD of the CM exhibits diffusive behavior at very short times ($t \lesssim 0.2 t_\omega  $) and, starting from 
$t \approx  4 t_\omega$,
a ballistic motion evidently due to the forward swimming of the \tb\ is observed. Interestingly, during the intermediate time span, the active \tb\ first triggers a superdiffusive increase in the MSD, owing to its forward swimming, but then slows down before becoming ballistic. The slow-down is due to the MSD of the relative motion, $\langle| \Delta \boldsymbol{R}_c(t) |^2\rangle$.
Initially, it shows a similar behavior as the CM, but then the MSD slows down since its value is obviously limited. At around $t \approx 2 t_\omega$, the MSD starts to oscillates with an average period of approximately $4.34 t_\omega$. 
The oscillations  originate from the relative vector $\Delta \r(t) = \check{\r}(t) - \check{\r}_c(t)$ rotating continuously around the centerline. For a perfectly straight helical path, the oscillations would continue. Here, they are damped since the local orientation vector $\boldsymbol{\tau}(t)$ of the helix varies in time, and the MSD approaches a constant value of ca.\ $3 D_0^2$, determined by the diameter $D_\text{helix}$ of the helical trajectory. In the MSD of the helical centerline, $\langle|\boldsymbol{R}_c(t) |^2\rangle$, the relative motion is not included and the MSD smoothly crosses over from a superdiffusive regime at small times into ballistic motion. Finally, 
an effective swimming velocity along the local helix axis
can be obtained from the MSD at long times using $v = \text{d}\sqrt{\langle|\boldsymbol{R}(t)|^2\rangle}/\text{d}t$. For the above \tb, the measured velocity is 
$0.0116v_\omega$. Note that the times considered here are still much smaller than the correlation time of the swimming direction, which then decorrelates due to thermal motion. Thus, the resulting long-time diffusive regime is not captured in  Fig.~\ref{fig:bulk}. Mapping to physical units, we 
	obtain a beating frequency for the in-silico \textit{T. brucei} of $f_0 = \omega/2\pi =38.6\text{Hz}$, which then gives a typical swimming speed of $v \approx 70.4\mu\text{m/s}$, which has the same magnitude as the experimentally observed swimming velocity of \textit{T. brucei} of approximately $30\mu\text{m/s}$.

\subsubsection{Diameter of helical trajectory: $D_{helix}$}
\begin{figure}
\centering
	\includegraphics[width=0.55\columnwidth,trim={0 0 0 0},clip]{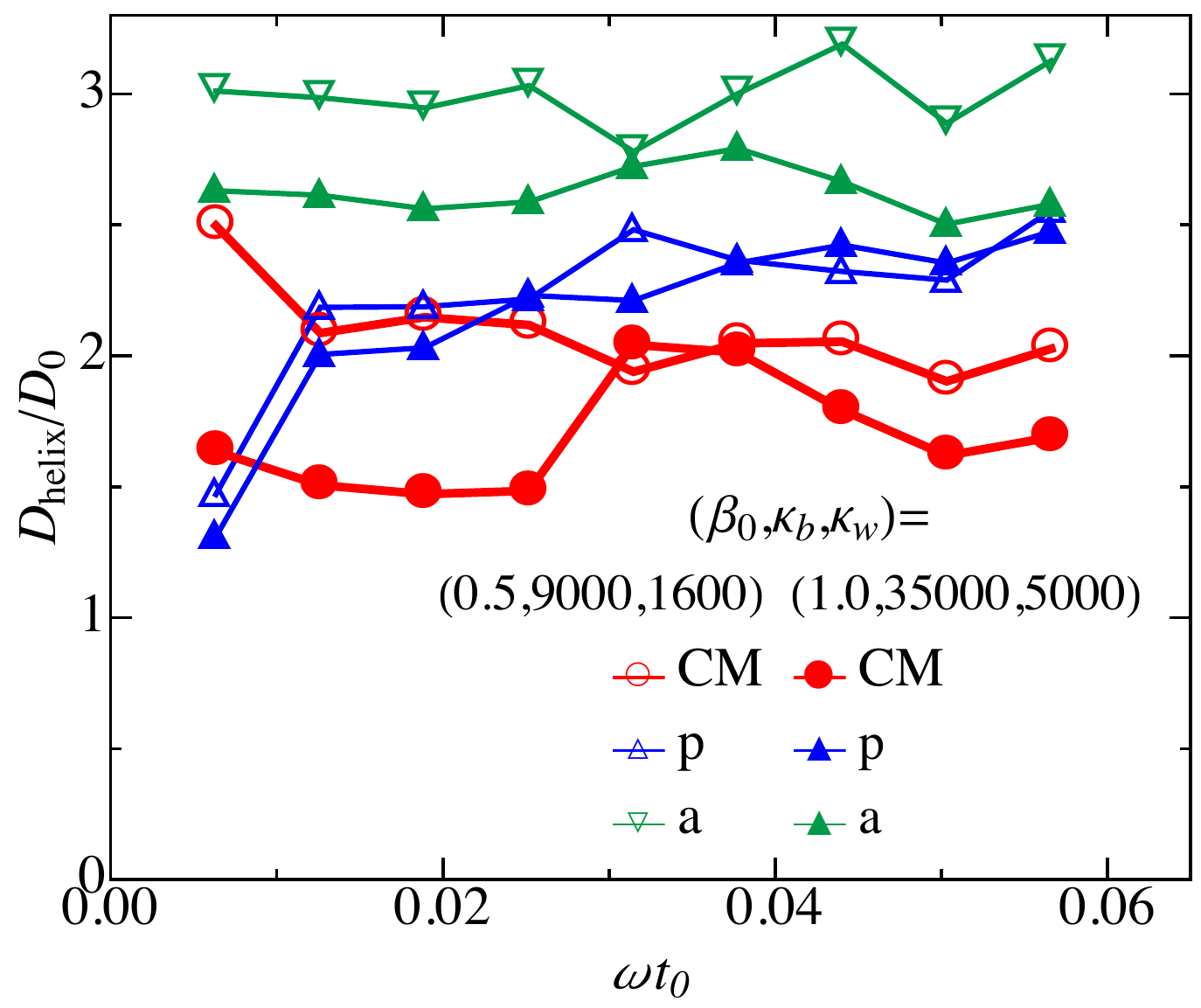}
\caption{\label{fig:dhelix}Average diameters of the helical trajectories for the center of mass (blue), anterior (red), and posterior (green) 
ends plotted \textit{vs.} the bending-wave angular frequency $\omega$. Results shown with filled symbols correspond to simulations 
with our standard set of parameters in table\ \ref{tb:parameters}. The open symbols refer to a case with smaller bending rigidities 
and smaller angular amplitude $\beta_0$ as indicated in the legend.
}
\end{figure}

The above discussion suggests that the helical nature of the trajectory is closely related to the swimming properties. Hence, we also scrutinize how the average helical diameter $D_\text{helix}$ depends on the applied mechanical parameters. Figure~\ref{fig:dhelix} displays the estimated diameters $D_\text{helix}$ of the CM, anterior, and posterior-end trajectories plotted \textit{vs.} the angular frequency $\omega$ of the flagellar bending wave. For the \tb\ modeled with the standard parameter set we use filled symbols and open symbols to refer to a set of smaller bending rigidities $\kappa_{b0}$, $\kappa_w$.
For the latter parameters we use a bending-wave amplitude of $\beta_0 = 0.5 / a_0$. Interestingly, the diameters $D_\text{helix}$ for the same parameter set are roughly independent of $\omega$. In particular, this applies to the thin anterior end. For the thick posterior end, we realize some decrease with decreasing $\omega$, while for the CM no clear trend is visible, especially for the standard parameters. 
\footnote{Note that the helical trajectory is not perfect, so the helical diameter varies locally and from one full helical turn to the other.
The plotted diameters are an average over helical paths with a length of $N=25-50$ pitches and the simplest estimate for 
the relative standard error is $1/\sqrt{N}$, which gives $14 - 20\%$.}
This rough independence from $\omega$ confirms that we are in the quasistatic limit of swimming, where the \tb\ goes through a sequence of equilibrium shapes during the flagellar beating, as we recognized earlier \cite{Alizadehrad:2015}. We also observe that in most cases the helix diameter decreases in the order of anterior end,
posterior end, and CM. 
Thus, the \tb\ performs some ``kayaking" motion around the CM during swimming and the cells have an overall ``crescent-like" twisted shape.
Figure\ \ref{fig:dhelix} shows that for the thin anterior end the helix diameter increases for the cell with smaller bending rigidities, which makes sense since at the anterior end we have the largest deformations. Otherwise, for the thick posterior end no variation is visible and for the CM only for $\omega t_0 < 0.03$.

We further examined the helix diameter for the center-of-mass trajectory in simulations by systematically varying our
mechanical parameters, whereas keeping the beating frequency fixed at $\omega t_0 = 0.038$, as reported in the four blocks of 
Table~\ref{tb:dhelix}. We also show the corresponding mean values for the
end-to-end distance $l_\text{e2e}$ between the anterior and posterior end of the swimming parasite given relative to the contour length $L_0$. The two upper blocks correspond to cases with the same flagellum bending rigidity, $\kappa_w = 5000 \, k_BT /a_0^2$, where the 
left block uses an amplitude of $\beta_{0} a_0=1$ and the right uses $\beta_{0} a_0=0.5$. 
Similarly, the two lower blocks have a reduced flagellum stiffness of $\kappa_w = 1600 \, k_BT /a_0^2$. In 
all four blocks
the longitudinal bending stiffness $\kappa_{b0}$ is varied from 
$10000$ to $35000k_BT$. Note, for $\kappa_{b0} = 10000k_BT$ and smaller values the helical trajectory becomes irregular and even segments occur that are only weakly bent. So the corresponding values for $D_\text{helix}$ and $l_\text{e2e}$ should be viewed with caution and we will not consider it further.
The first insight is that for bending rigidities $\kappa_{b0}$ above $10000 \,k_BT$, the diameter
$D_\text{helix}$ remains almost constant, and the corresponding values of $l_\text{e2e}$ vary by less than $5\%$. 
For the blocks with $\kappa_w = 5000 \, k_BT / a_0^2$, a reduced bending wave amplitude $\beta_{0} a_0$ leads to a smaller $D_\text{helix}$. This is somewhat intuitive, as the bending deformation exerted by the bending wave is weaker, which is documented by the increased end-to-end distance $l_\text{e2e}$. For $\kappa_w = 1600 \, k_BT / a_0^2$ and $\beta_{0} a_0 = 1.0$ (lower left block), we again observe that a reduced forcing
by the flagellar wave increases $l_\text{e2e}$ and $D_\text{helix}$ becomes smaller. However, in contrast to the upper blocks, for the lower blocks with the reduced $\kappa_w$ the smaller amplitude $\beta_{0} a_0$ does not significantly alter $D_\text{helix}$, while $l_\text{e2e}$ increases according to our expectations. Therefore, we examined the helix diameters of the anterior and posterior ends more closely and noticed that here the helix diameters are clearly reduced. These results indicate that the movement of the model parasite highly depends on the combination of mechanical parameters chosen. This also demonstrates a broad range of tunability for the model.

\begin{table} 
\centering
\caption{Mean helix diameter $D_\text{helix}$ of the center-of-mass trajectory and the normalized end-to-end distance $l_{e2e}/L_0$
of the swimming parasite for various parameters: the longitudinal bending rigidity $\kappa_{b0}$, two flagellar bending rigidities
$\kappa_w$ (upper and lower blocks), and two wave amplitudes $\beta_0a_0$ (left and right blocks). The beating frequency is always $\omega t_0 = 0.038$.
}
\renewcommand{\arraystretch}{1.5}
\begin{tabular}{|c|| c | c | c | c |c||c | c | c | c |c|  }
\hline
$\kappa_w = 5000 \,  k_BT /a_0^2$
& \multicolumn{5}{c||}{$\beta_{0} a_0 = 1.0$} & \multicolumn{5}{c|}{$\beta_{0} a_0 = 0.5$} \\
\hline
$\kappa_{b0} \,\, / 10^4 k_BT $  & 	 
$1$ & $1.5$  &	$2$&	$3$&	$3.5$ & 	 
$1$ & $1.5$ & $2$&	$3$ & $3.5$   \\
\hline
$D_\text{helix}/D_0 $& 	 
$2.26$ & $2.00$	& $1.98$ &	$1.91$ &	$2.02$  &	
$1.90$ & $1.52$ & $1.46$ &	$1.37$ &	$1.47$ \\
\hline
$l_{e2e}/L_0$& 	 
$0.58$ & $0.64$ & $0.66$ &	$0.69$ &	$0.70$  &	
$0.83$ & $0.83$ &  $0.83$ &	$0.85$ &	$0.86$ \\
\hline \hline
$\kappa_w = 1600 \, k_BT /a_0^2$
& \multicolumn{5}{c||}{$\beta_{0} a_0 = 1.0$} & \multicolumn{5}{c|}{$\beta_{0} a_0 = 0.5$} \\
\hline
$\kappa_{b0} \,\, /10^4 k_BT$ & 	 
$1$ & $1.5$  & $2$&	$3$&	$3.5$  & 	 
$1$ & $1.5$ & $2$ &	$3$ & $3.5$  \\
\hline
$D_\text{helix}/D_0 $& 	 
$1.99$ & $1.74$   & $1.71$ &	$1.74$ &	$1.69$ & 	 
$1.62$ & $1.72$	& $1.74$ &	$1.77$ &	$1.67$  \\
\hline
$l_{e2e} $& 	 
$0.72$ & $0.74$& $0.75$ &	$0.78$ &	$0.80$  &	
 $0.88$ & $0.89$ & $0.90$ &	$0.92$ &	$0.92$ \\
\hline
\end{tabular}
\label{tb:dhelix}
\end{table}

\subsubsection{Rolling motion of the cell body}
Due to the chiral attachment of the flagellum, the cell performs rotational motion around its long axis resembling a ``log-rolling" motion. Previous experimental work~\cite{Heddergott:2012} and simulation studies~\cite{Alizadehrad:2015} reported that the beating frequency $\omega$ is approximately $8 \pm 2$ times faster than the rolling frequency $\Omega$. This means a full rotation of the cell body about its long axis takes  $8 \pm 2$ flagellar beat cycles. This value has been used as an important parameter to map the \textit{in silico} \tb\ to experimental observations~\cite{Alizadehrad:2015}.

To rigorously quantify the rolling motion, we adopt an approach that measures the accumulative rolling angle about the local centerline of the cell body between the 5th and 6th segment. Since the cell body itself changes orientation in space, 
we monitor the rolling angle in time steps $\Delta t \ll t_\omega$, where the cell body hardly deforms and utilize a gradient descent optimization to compute the rolling angle increment. A more detailed explanation of this method is provided in Appendix~\ref{sec:apps}. Using the simulation setup with our standard parameters shown in Table~\ref{tb:parameters}, the \tb\ achieves the ratio $\omega/\Omega \approx 7$, which aligns well with the experimentally observed value of $8\pm2$~\cite{Heddergott:2012}.

Based on our analysis so far, we use the parameters listed in Table~\ref{tb:parameters} here and in subsequent sections to study the locomotion of \tb\ in microchannels.

\subsection{Straight cylindrical channel}
\label{subsec:cylinder}

When swimming in a confined straight cylindrical channel, interactions with the curved bounding surface are expected to influence the behavior of \tb. Given that the typical length of a \tb\ ($25~\mu\mathrm{m}$) in its bloodstream form is approximately two to four times larger than the average mammalian capillary diameter ($5\textendash10~\mu\mathrm{m}$)\cite{Li:2009}, the parasite will be directed to swim
along the capillary~\cite{Doro:2019}, particularly in the presence of blood flow. Accordingly, in our simulations, the \tb\ is initially positioned at the center of the channel with its long axis aligned parallel to the channel. Notably, the channel appears to ``guide" the swimming of the \tb\ to some extent, especially when the channel width is comparable to the helix diameter observed in bulk (see supplementary Video 1).
\begin{figure}
\centering
\includegraphics[width=0.55\columnwidth,trim={0 0 0 0},clip]{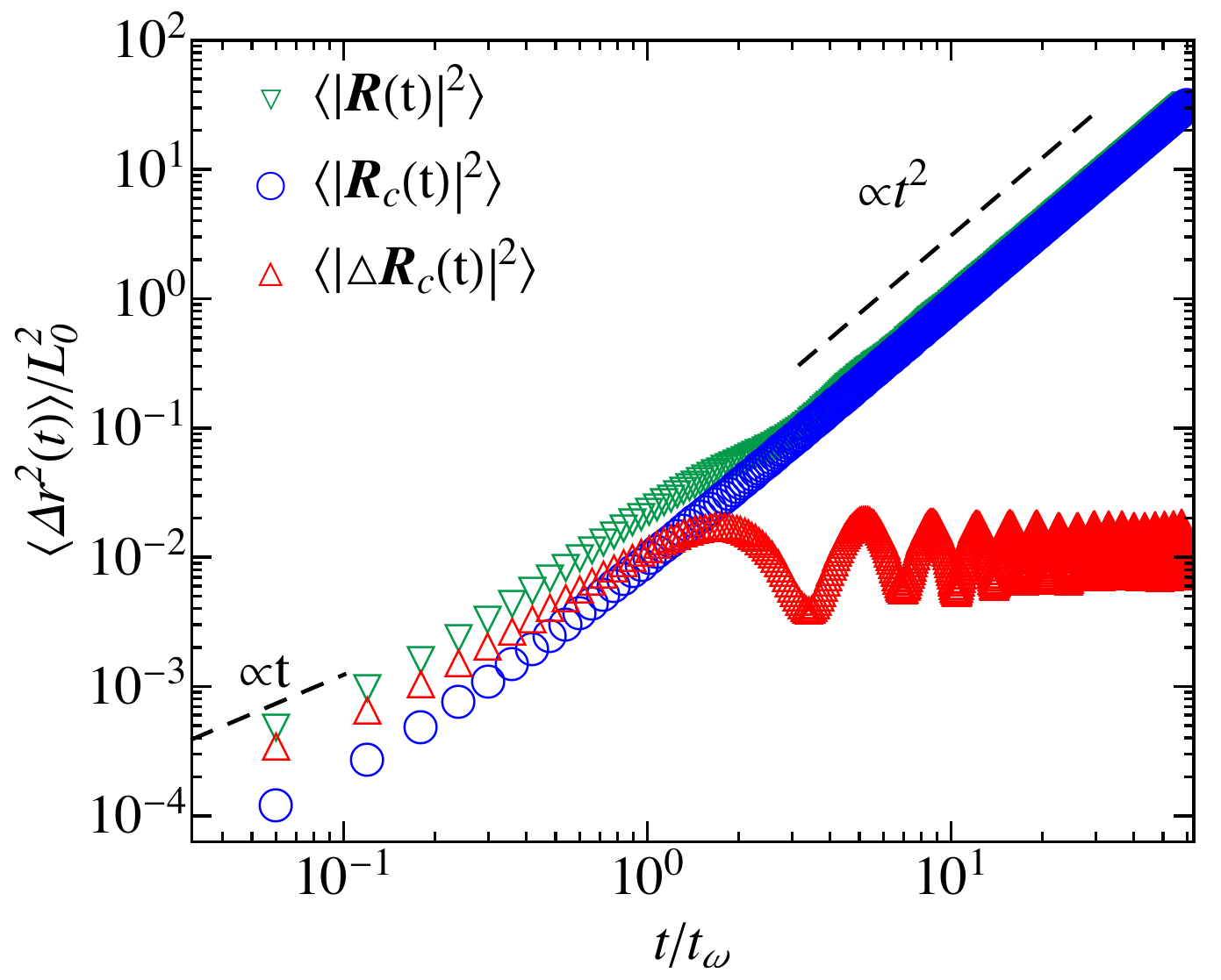}
\caption{\label{fig:chmsd}Time evolution of the MSDs for the CM $\langle|\boldsymbol{R}(t) |^2\rangle$, the helix centerline $\langle|\boldsymbol{R}_c(t) |^2\rangle$, and the relative motion $\langle| \Delta \boldsymbol{R}_c(t) |^2\rangle$ of a \tb\ swimming in a cylindrical channel with channel width $D_\text{ch}=4.66D_{0}$. Symbols and colors are chosen as in Fig.~\ref{fig:bulk}(c).
}
\end{figure}

\begin{figure} 
\centering
\includegraphics[width=0.55\columnwidth,trim={0 0 0 0},clip]{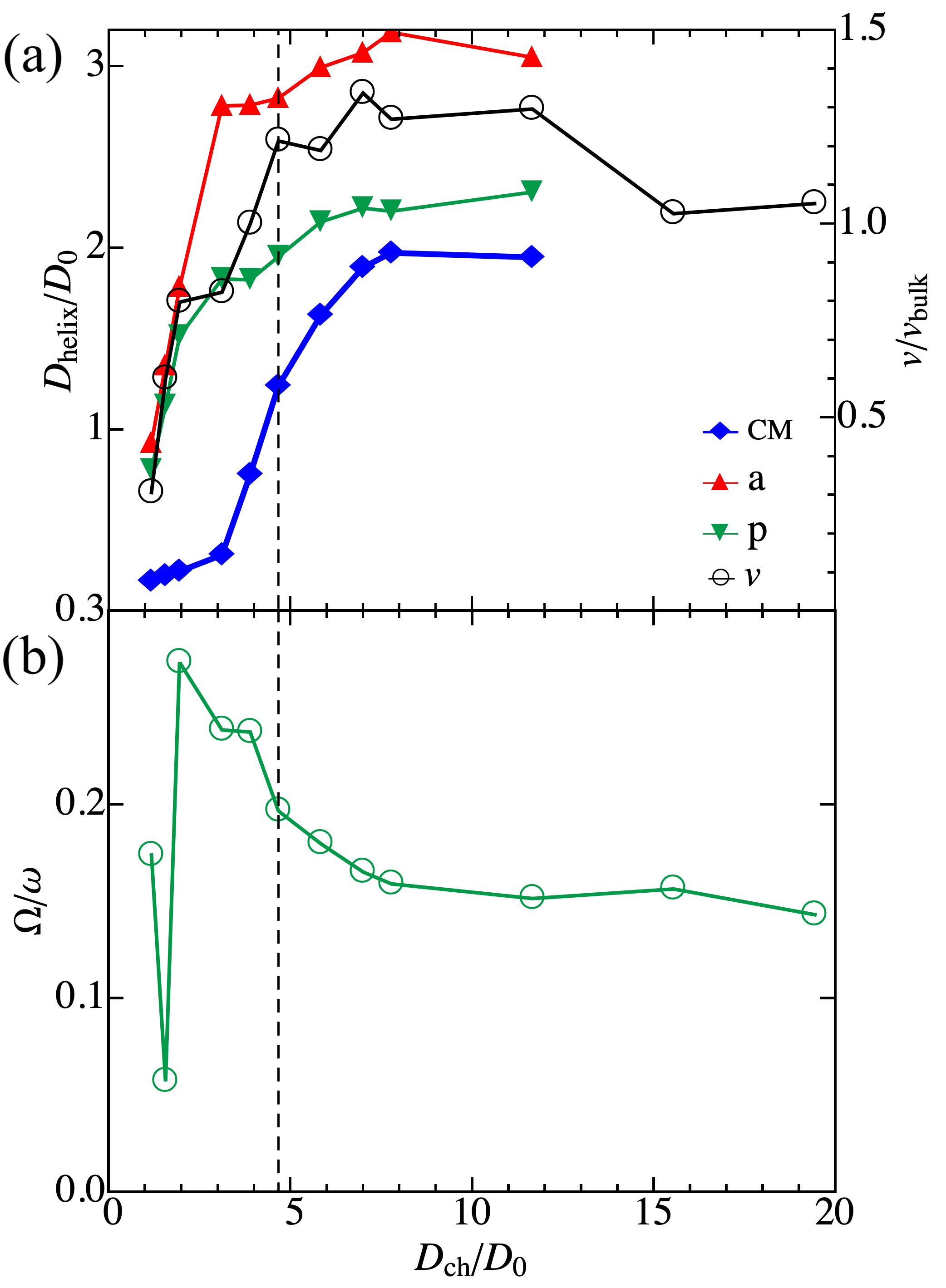}
\caption{\label{fig:ch}(a)~The average helix diameters for the trajectories of the CM (blue filled diamonds), anterior (red upper triangles) and posterior end (green lower triangles), as well as the swimming speed $v$ (open circles, normalized by the bulk value $v_{\text{bulk}}$), plotted \textit{vs.} the cylindrical channel width $D_\text{ch}$. (b)~The frequency $\Omega$ of body rolling as a function of channel width $D_\text{ch}$ for a \tb\ swimming in straight cylindrical channels. The dashed vertical line marks the channel simulations corresponding to Fig.~\ref{fig:chmsd}.
 
}
\end{figure}

Figure~\ref{fig:chmsd} displays the MSD results of the CM $\langle|\boldsymbol{R}(t) |^2\rangle$, the helix centerline
$\langle|\boldsymbol{R}_c(t) |^2\rangle$, and the relative motion $\langle| \Delta \boldsymbol{R}_c(t) |^2\rangle$ of a \tb\ swimming in 
a cylindrical channel with a channel width $D_{\text{ch}}=4.66D_{0}$. The symbols and colors used are consistent with those in Fig.\ \ref{fig:bulk}(c).
Interestingly, these curves largely resemble the results obtained from the bulk simulations. However, upon closer inspection, the curve for $\langle| \Delta \boldsymbol{R}_c(t) |^2\rangle$ shows the aforementioned periodic oscillations around a mean value of $1.2D_0^2$
but now with a constant amplitude, for times beyond $t_\omega$. This makes sense since the local axis of the helical trajectory always points approximately along the channel axis. Furthermore, in contrast to the \tb\ swimming in bulk, a smaller oscillation period of $3.52t_\omega$ is observed, also suggesting that the swimming trajectory inside the cylindrical channel is rectified into a straighter helical shape, now with a smaller diameter.
Using the method described in Sec.\ \ref{subsec:bulk}, the average diameter of the helix for the CM is estimated 
to be $D_{\text{helix}}=1.24D_0$ (cf. corresponding data point in Fig.~\ref{fig:ch}(a) marked by the dashed vertical line). Indeed, this value is smaller than $D_{\text{helix}}= 2.02 D_0$ observed in bulk. Furthermore, the channel size $D_\text{ch}$ is also wider than the larger helix diameter of the anterior end, for which we determined $2.8D_0$ in bulk for our standard parameters (see Fig.~\ref{fig:dhelix}). Thus, the \tb\ has already some mechanical adaptation with respect to its helical trajectory, which we attribute to the fluid-mediated long-range hydrodynamic interactions with the channel wall.

To further illustrate the effect of the channel on the helical path, we plot in Fig.\ \ref{fig:ch}(a) the helical diameter $D_\text{helix}$ for the CM, anterior, and posterior end as a function of channel width $D_\text{ch}$. Typically, the helix of the anterior tip has the largest diameter among the three characteristic cell-body points. As the channel narrows, the average helical diameters remain close to their bulk values until $D_{ch}\approx 7D_0$, after which a moderate reduction in $D_\text{helix}$ is observed until $D_\text{ch}\approx 3D_0$. Only $D_\text{helix}$ for the CM declines sharply showing that  the helix becomes aligned within the channel. In this region, the difference $D_\text{ch}-D_{\text{helix}}$ is always larger than $D_0$, especially also for the anterior tip with the largest diameter, indicating that the cell body is preferentially kept away
from the wall surface, owing to large deformations of the cell body and hydrodynamic interactions. When $D_\text{ch}\approx 3D_0$, the 
diameters of the anterior and posterior tip sharply drop. At this point, the beating anterior tip almost touches the channel wall. The 
corresponding diameter gradually approaches that of the posterior end until at $D_\text{ch}\approx 3D_0$, where they are nearly the same. 
This indicates that the \tb\ lengthens in tighter confinements. 

Confinement not only governs the helical diameter, $D_\text{helix}$, but also the corresponding swimming velocity $v$, as both show 
similar trends with varying channel widths. Interestingly, a slight enhancement in velocity by $35\%$ relative to the bulk value 
$v_\text{bulk}$ is observed for decreasing $D_\text{ch}$ with a broad maximum at $D_\text{ch}\approx 7D_0$, coinciding with the
 onset of the suppression of $D_\text{helix}$. This enhancement arises from the ability of microswimmers to exploit no-slip boundaries 
 to increase their swimming speed. Furthermore, the characteristic helical trajectory of \tb\ generates a rotating flow, which circulates in 
 the cylindrical geometry to facilitate movement.
Note, for a straight attachment of the flagellum, where the rotating flow does not exist, we find that the swimming velocity is
 smaller both in the channel and in the bulk fluid in agreement with our previous study in Ref.\ \cite{Alizadehrad:2015}.
However, as the channel narrows further, the \tb\ loses the advantage of the strongly  expansive stroke of the anterior tip, resulting in 
a decrease in swimming velocity. Notably, for the narrowest channel studied ($D_\text{ch}=1.17D_0$), a speed of $0.3v_{\text{bulk}}$ 
is still maintained. 
Interestingly, a similar curve for swimming speed versus channel width has recently been observed for \textit{E.\ coli} 
 moving in a microchannel \cite{Vizsnyiczai:2020}. There, also a  broad maximum with a $10\%$ enhancement of the velocity is
 recorded. Notably, the optimum swimming speed occurs when the bacterium transitions from swimming along the channel wall to 
 axial swimming along the centerline. The location of the transition and the maximum is determined by the wobbling amplitude of the cell.
This is similar to our case, where the maximum occurs when the helix  becomes aligned along the channel axis.


It is also instructive to examine how the ``log-rolling" angular frequency $\Omega$ alters within the microchannel. As indicated in 
Fig.~\ref{fig:ch}(b), $\Omega$ increases as the channel narrows and reaches a peak at approximately $D_\text{ch}=2D_0$ with nearly 
twice the bulk rolling frequency. This enhanced rolling motion should be also due to the rotating flow generated by the swimmer and 
by which it interacts with the no-slip channel walls. However, as the channel becomes tighter, $\Omega$ drastically slows down.
The supplementary Video\ 1b for $D_\text{ch}=1.55D_0$, where the minimum rotational frequency occurs, shows that the wobbling 
motion of the whole cell body is hindered by the contact with the channel walls and thereby $\Omega$ decreases, also due 
to the increased friction.
However, surprisingly at $D_\text{ch}=1.17D_0$, $\Omega$ reaches a value even slightly higher than 
the bulk rolling frequency. 
Here, the cell body is straightened, the wobbling motion has ceased, and the trypanosome as a more straight object can more easily 
rotate about its long axis.
%

\subsection{Cylindrical channel with constrictions}
\label{subsec:constriction}
We take a step further to explore how the \tb\ interacts with and navigates through a constriction inside a microchannel.
When crossing the  blood-brain barrier (BBB) to infect the brain, the \tb\ has to squeeze through tight passages \cite{Abbott:2010,Pays23} and our study might contribute to an understanding of this process.
Using the microchannel with the constriction designed as shown in Fig.~\ref{fig:cone}, we investigate if and how the \textit{in silico} \tb\ swims through such a constriction. As before, the \tb\ is initially positioned on the channel axis close to the left end and oriented along the axis so that it swims towards the constriction. Depending on the length and width of the constriction, we observe three types of trajectories: the trypanosome either slips through the constriction, performs a  stuck-slip motion, or becomes stuck as illustrated by the three videos in the supplemental material. We first describe the first two types of trajectories in detail, quantify the retention time in the constriction, and finally show a state diagram in the parameter space of the constriction, diameter $D_n$ \textit{vs.} length $L_n$.

\begin{figure*}
\centering
\includegraphics[width=0.48\columnwidth,trim={0 0 0 0},clip]{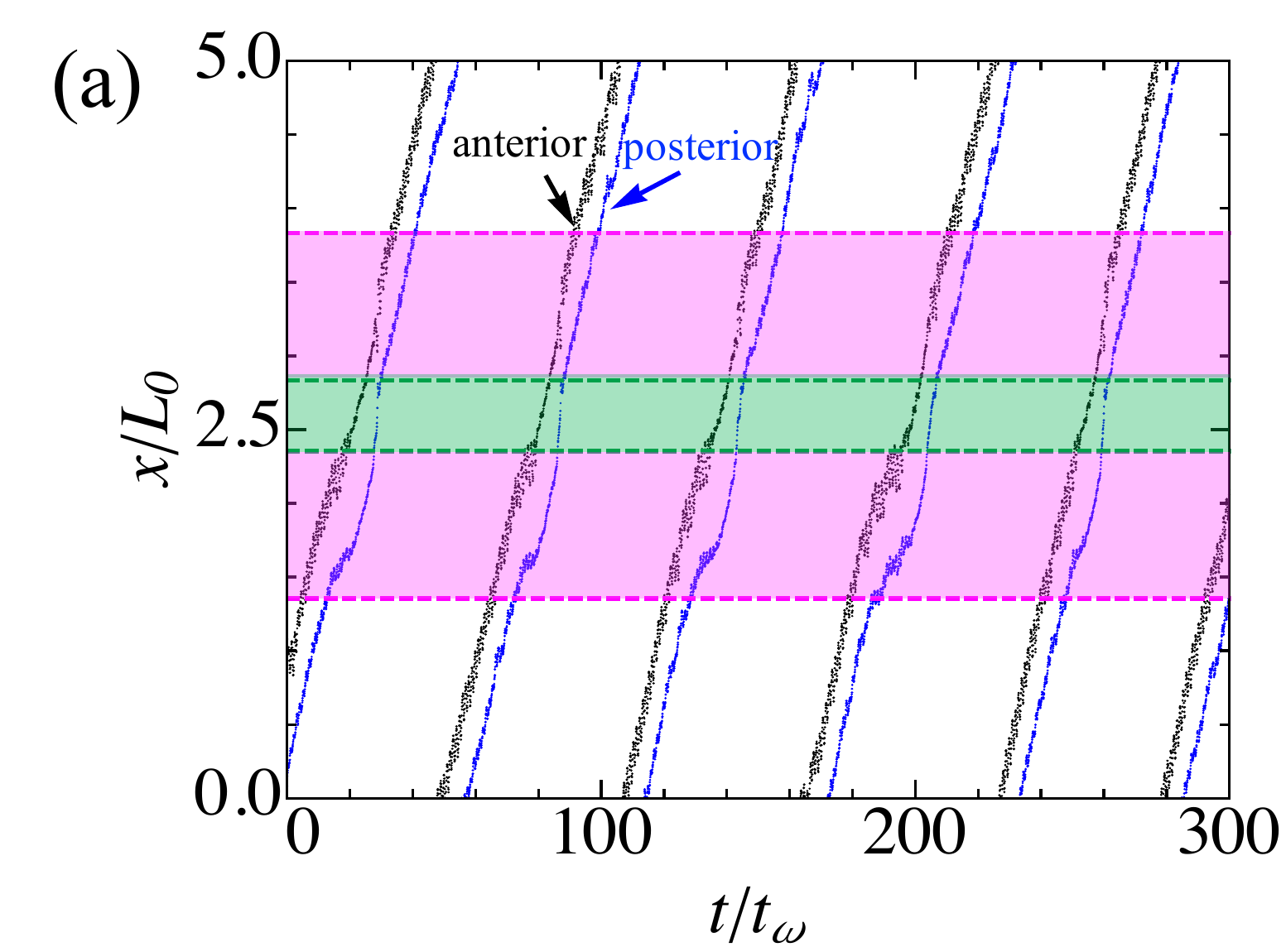}
%
\centering\includegraphics[width=0.51\columnwidth,trim={0 0 0 0},clip]{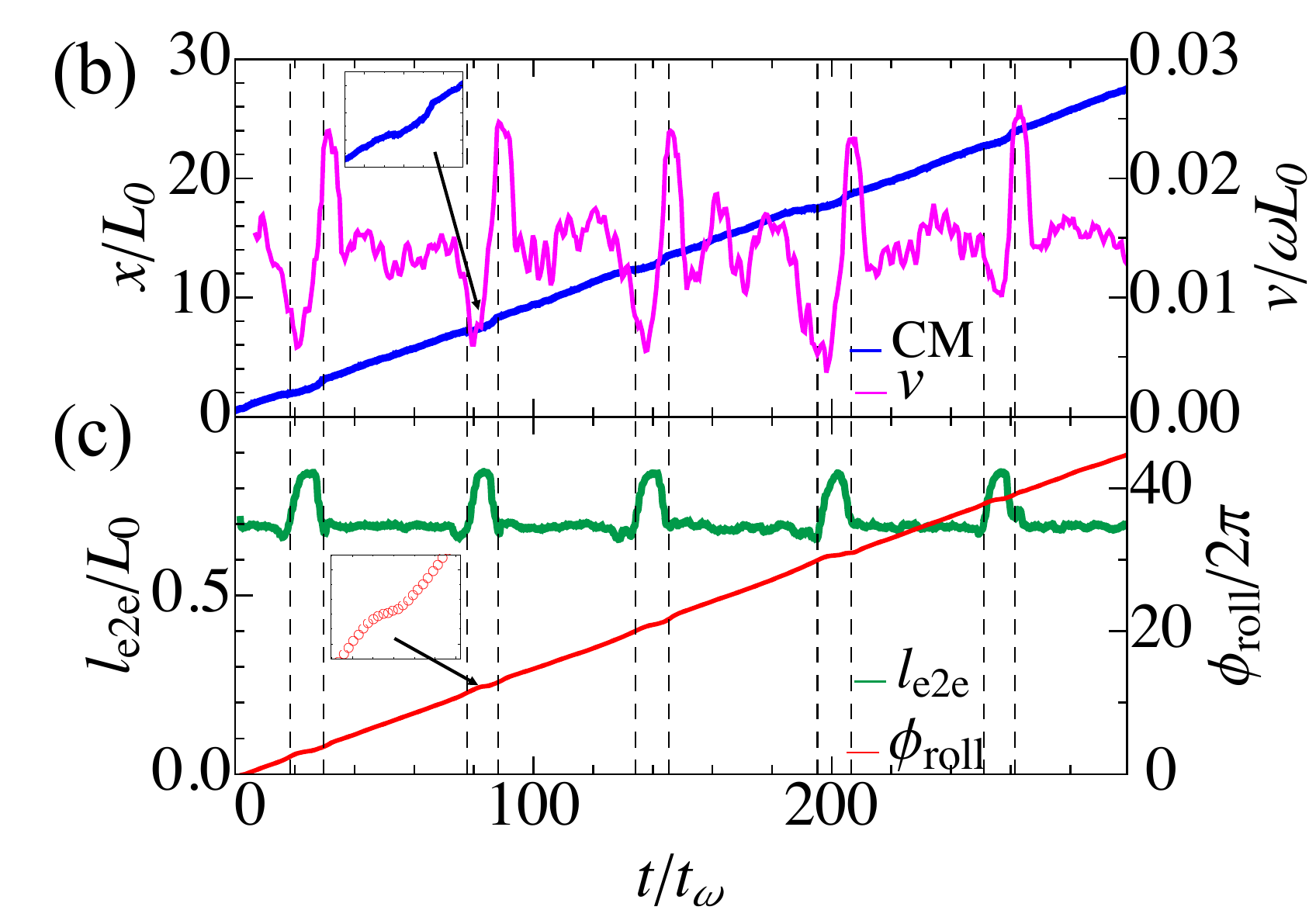}
\caption{\label{fig:det_ret} 
\tb\ swimming in a channel through a constriction of width $D_n=1.4D_0$ and lengh  $L_n=0.48L_0$.
(a) The $x$ position along the channel axis of the anterior (black) and posterior (blue) ends plotted \textit{vs} time. The repeated
trajectories result from the PBCs. The magenta and green shaded areas represent the cone-shaped and constriction regions of 
the microchannel, respectively.	 
(b) The $x$ position of the CM (blue) and its instantaneous velocity $v_x$ (magenta) along the $x$-direction plotted \textit{vs.}
time. An inset highlights the trajectory in the constriction.
(c) The end-to-end distance $l_\text{e2e}$ and rolling angle $\phi_\text{roll}$ plotted \textit{vs.} time.
The five pairs of dashed lines indicate the times when the CM enters and exits the constriction,
five times. An inset highlights the rolling angle in the constriction.
}
\end{figure*}

\subsubsection{Slip motion}
\label{sec:slip}

Figure~\ref{fig:det_ret}(a) shows typical trajectories of the anterior (black dots) and posterior (blue dots) ends of the \tb\ obtained from an MPCD simulation with a constriction of diameter $D_n=1.4D_0$ and length $L_n=0.48L_0$. The trajectories crossing several times the channel of length $5L_0$ result from the periodic boundary conditions (PBCs) applied along the channel. The constriction is highlighted in green, while the cone-shaped regions are marked in magenta. The simulation clearly illustrates how the \tb\ traverses the cone region (length $0.962L_0$, as depicted in Fig.~\ref{fig:cone}) and begins probing the narrow constriction with its anterior tip. This probing behavior is evident from the larger clustering of dots just before the \tb\ enters the constriction and a clear slow down of the posterior end (cf. supplementary Video 2). Due to the large excursions of the anterior tip exceeding the constriction width, the tip initially collides with the tilted wall of the cone, resulting in a momentary slow-down in swimming velocity, before the trypanosome slides straight through the constriction. Upon exiting, particularly when the anterior tip sticks out while the rear part remains inside the constriction, the channel expands for the anterior tip, which again starts to beat with large excursions. As we discuss below, this helps the \tb\ to exit the constriction.

Accordingly, the CM position projected on the channel axis with unwrapped PBCs is shown as $x(t)$ by the blue curve in Fig.~\ref{fig:det_ret}(b). The five pairs of dashed vertical lines indicate the times when the CM enters and exits the constriction during the simulation. The trajectory is relatively smooth through the wider channel and cone regions but exhibits a noticeable dent, which starts when the anterior tip tries to enter the constriction (see the close-up segment in the inset). The corresponding instantaneous velocity $v_x(t)$ is depicted by the magenta curve in Fig.\ \ref{fig:det_ret}(b). It varies around $0.016 \, \omega L_0$, when the trypanosome swims in the wider channel parts, but slows down to approximately $0.008 \, \omega L_0$ starting just before the trypaosome enters the constriction. This is consistent with the observation of the dent in the trajectory. Interestingly, the \tb\ speeds up to a velocity of $0.024\, \omega L_0$ while exiting, which we will address below.

In comparison, the rolling motion of the \tb, depicted by the rolling angle $\phi_\text{roll}$ in Fig.\ \ref{fig:det_ret}(c) (red curve), exhibits a deceleration but no corresponding acceleration within the constriction, as shown by the close-up in the inset. Moreover, the end-to-end distance $l_\text{e2e}$, represented by the green curve, reveals further details. In the wider cylindrical and cone-shaped regions of the channel, where the \tb\ has sufficient space for its beating anterior end, $l_\text{e2e}$ remains at the bulk-swimming value of $0.70 L_0$. However, within the constriction, the cell body of the \tb\ is straightened and $l_\text{e2e}$ increases to $0.85L_0$. The arising peaks in the time evolution of $l_\text{e2e}$ show a slightly skewed shape, owing to the different speeds and deformations of the \tb\ when entering and leaving the constrictions. 
By analyzing the four properties presented in Fig.~\ref{fig:det_ret}, we conclude that the \tb\ decelerates before entering the constriction but accelerates upon exiting. This behavior is accompanied by distinctive changes in the end-to-end distance reflecting adjustments in the \tb's body configuration before, during, and after its passage through the constriction.

\begin{figure}
\centering\includegraphics[width = 1.0 \columnwidth,trim={0 0 0 0},clip]{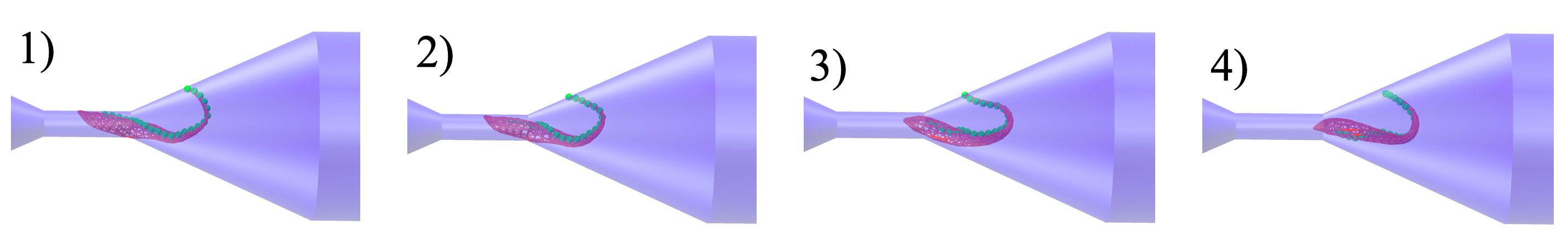}
\caption{\label{fig:snapshots} Snapshots from supplementary Video 2
for the slip motion to illustrate the acceleration when the \tb\ exits the constriction. They show how the anterior tip pushes against the wall while the bending wave travels 
along the cell body. The time interval between
the snapshots is
always ca.\ $0.12 t_\omega$. The snapshots are chosen between the $57^\text{th}$ and $58^\text{th}$ second in the supplementary Video 2.
}
\end{figure}

At first glance, it may seem counter-intuitive that the \tb\ speeds up beyond the bulk-velocity value when leaving the constriction. To understand this observation we carefully examine its movements. We observe that when the anterior end has left the constriction and performs again its large excursions, it utilizes the tilted wall in the cone-shaped area to establish a pivot to ``pull" the cell body out of the constriction. In other words, the anterior tip pushes against the wall, while the bending wave initiated by the beating flagellum propagates along the body and pulls the body out of the constriction. Then, the tip leaves the wall and starts a new stroke. This mechanism is illustrated by the snapshots in Fig.~\ref{fig:snapshots} and the supplementary Video 2. The reported behavior contrasts with the scenario before entering the constriction, where the anterior end, while pushing against the tilted wall, slows down the trypanosome, as discussed earlier (cf. supplementary Video 2).

\subsubsection{Stuck-slip motion}
\label{sec:stuckslip}

\begin{figure*}
\centering
\includegraphics[width=0.45\columnwidth,trim={0 0 0 0},clip]{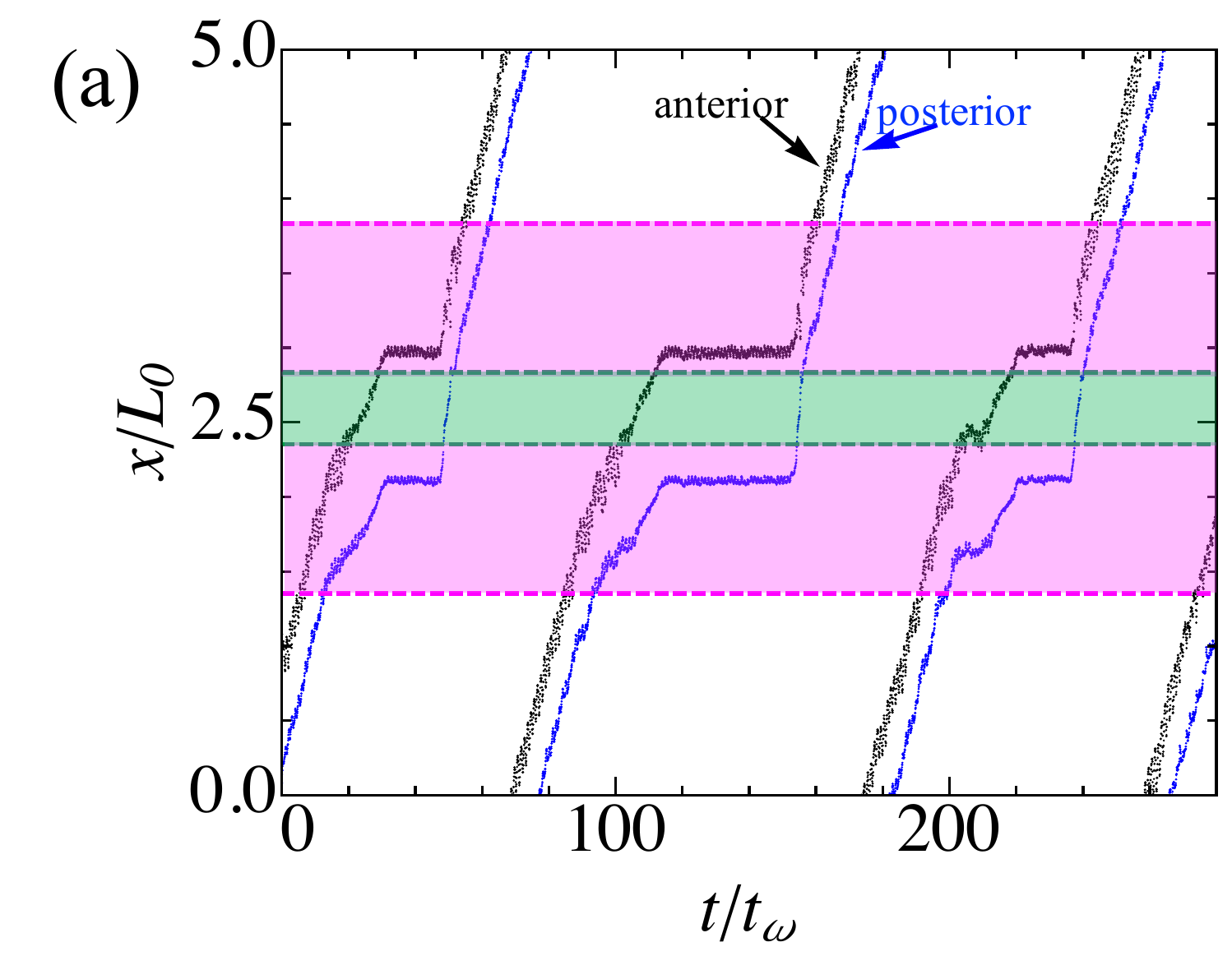}
\centering\includegraphics[width=0.5\columnwidth,trim={0 0 0 0},clip]{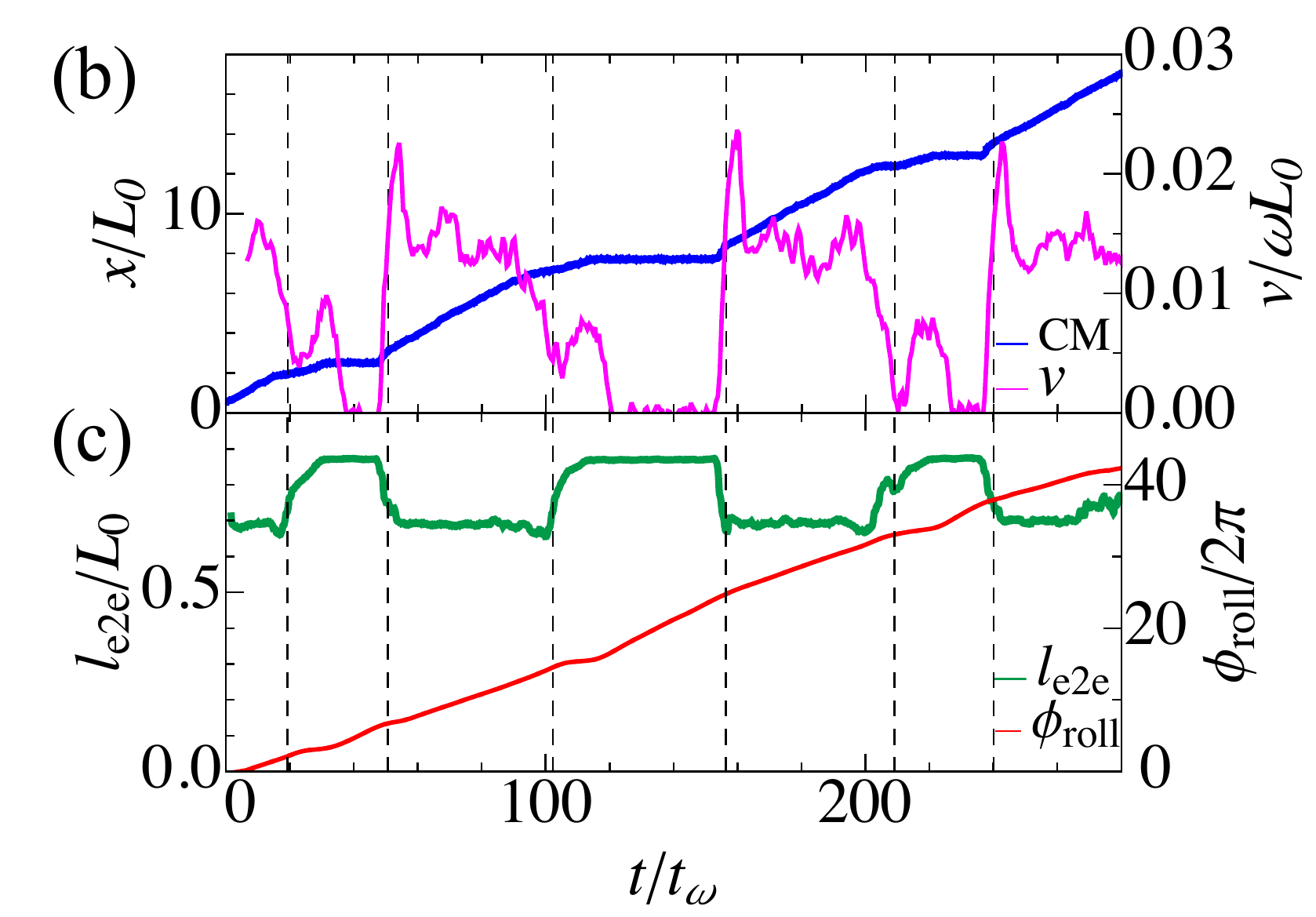}

\caption{\label{fig:stuck-slip} 
\tb\ swimming in a channel through a constriction of width $D_n=1.17D_0$ and lengh  $L_n=0.48L_0$.
(a) The $x$ position along the channel axis of the anterior and posterior ends plotted \textit{vs} time. 
(b) The $x$ position of the CM and its instantaneous velocity $v_x$ plotted \textit{vs.} time. (c) The end-to-end distance $l_\text{e2e}$ and rolling angle $\phi_\text{roll}$ plotted \textit{vs.} time. The three pairs of dashed lines mark the times when the CM enters and exits the constriction. The colors and symbols are consistent with those in Fig.~\ref{fig:det_ret}.
}
\end{figure*}

When narrowing the constriction width to $D_n=1.17D_0$ while keeping $L_n$ the same, we observe a distinctive swimming pattern compared to that observed during \textit{slip motion}. The occuring stuck-slip motion is displayed in Fig.\ \ref{fig:stuck-slip} and visualized in supplementary Video 3. The horizontal regions in the trajectories shown in Fig.\ \ref{fig:stuck-slip}(a) demonstrate, the tight space impedes the swimming within the constriction, but ultimately the \tb\ escapes. As before, the \tb\ decelerates before entering the constriction while the anterior end explores the space in front and then slips through it. Both the anterior and posterior ends exhibit a plateau in their trajectories while they are outside the constriction, but the intermediate body part becomes temporarily ``stuck". After a certain period of time, the anterior tip gets in contact with the wall of the cone zone and the rear body is pulled into and through the constriction. 

The features of the stuck-slip motion are clearly revealed by the CM trajectory and the instantaneous velocity $v_x$ shown in Fig.\ \ref{fig:stuck-slip}(b). Before the anterior end dips into the constriction, $v_x$ drops drastically from the bulk value at approximately $0.016\, \omega L_0$, and the cell shape becomes slightly more bent due to interactions with the tilted wall in the cone zone. This is visible in a small dip in the end-to-end distance $l_\text{e2e}$ in Fig.\ \ref{fig:stuck-slip}(c) just before the CM enters the constriction. Subsequently, the \tb\ adjusts its shape to better align with the constriction, leading to a rise in $v_x$ and $l_\text{e2e}$. However, as the \tb\ continues to squeeze through, the deforming \textit{thicker} posterior body part is hindered by the ``cone-to-constriction" region. 
The \tb\ stops moving forward (i.e., $v_x\approx0$) and the elongation of the cell body becomes the largest with $l_\text{e2e}=0.88L_0$. The CM entering the constriction is accompanied by a deceleration in its rolling motion as also indicated in Fig.~\ref{fig:stuck-slip}(c).

Further decreasing the constriction width, the trypanosome after several tries enters the constriction and ultimately becomes stuck 
there so that it cannot cross anymore. This is illustrated in the supplementary Video 4.

\subsubsection{Retention time and state diagram}
In the following, we systematically investigate the motion of the \tb\ in constrictions by varying $L_n$ and $D_n$. The obtained results 
are discussed in terms of \textit{retention time} and by constructing a \textit{state diagram} for the three motional states.

\begin{figure}
\centering
       \includegraphics[width=0.55\columnwidth,trim={0 0 0 0},clip]{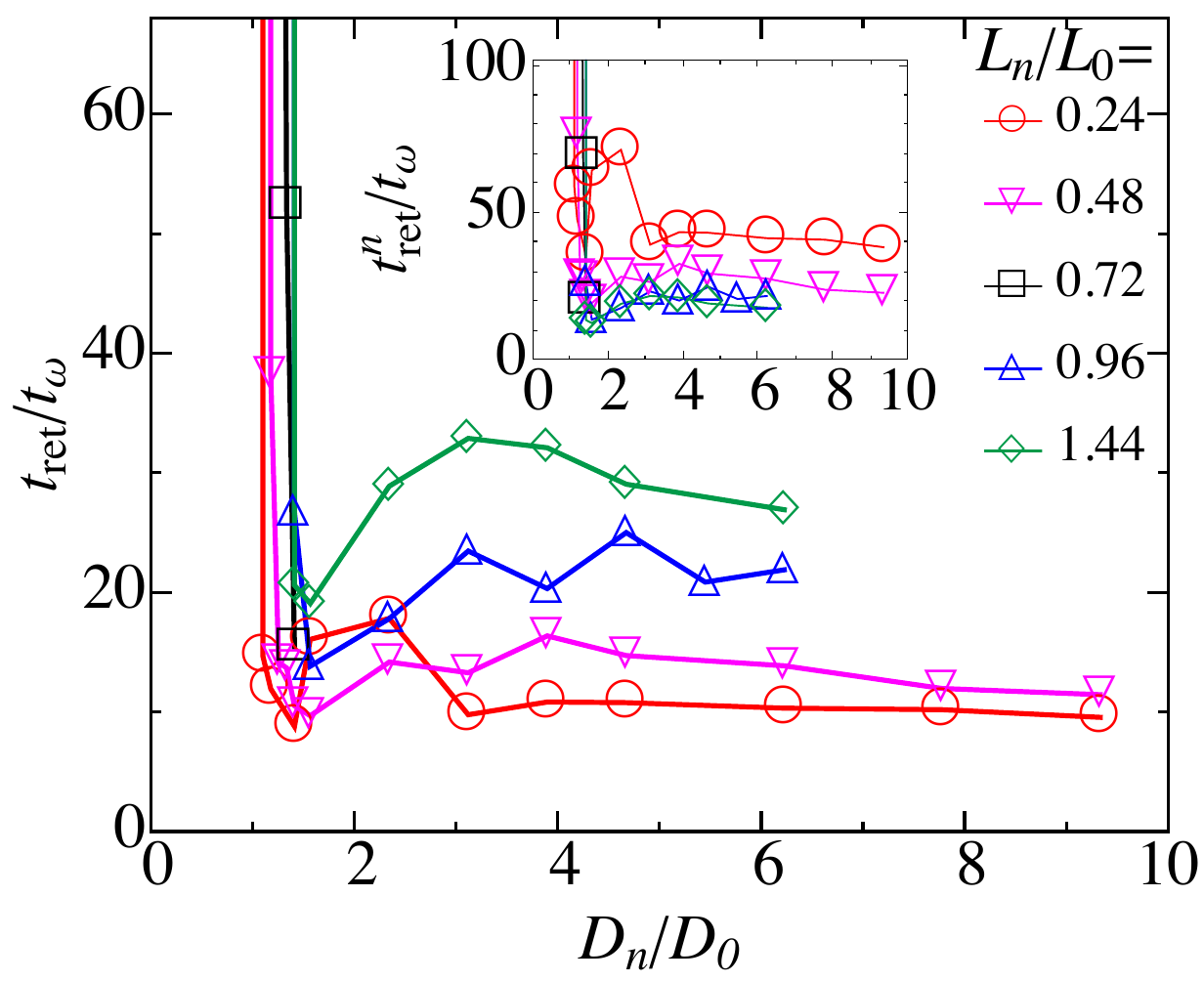}
%
%
\caption{\label{fig:ret} Retention time $t_\text{ret}$ \textit{vs.} constriction width $D_n$ for different lengths $L_n$.The inset displays the normalized retention time $t^n_\text{ret}$ (see main text for details).
}
\end{figure}

\paragraph{Retention time}
A natural question arises: how much time does the parasite spend, on average, inside the constriction before exiting? This is a quantity that is particularly relevant for biological and clinical studies on how the \tb\ crosses 
tight junctions like the blood-brain barrier. We define this duration as the ``retention time" $t_\text{ret}$ and investigate how it is affected by $D_n$ and $L_n$. Specifically, $t_\text{ret}$ is determined as the average time difference between the moment the anterior tip enters the constriction and the posterior tip exits. The results for the retention time as a function of constriction width $D_n$ for several lengths $L_n$ are presented in Fig.~\ref{fig:ret}. An overall trend is that $t_\text{ret}$ varies only slightly for $D_n \gtrapprox 3 D_0$ for the same $L_n$. Thus, the constriction does not have a noticeable influence on the swimming velocity of the \tb. When the constriction narrows, $t_\text{ret}$ slightly decreases and exhibits a minimum at $D_n \approx 1.4D_0$. This can be associated 
with the speed-up mechanism when leaving tighter constrictions, as discussed earlier.
While the anterior tip pushes against the wall, the bending wave pulls the cell body out of the constriction.
Ultimately, the retention time diverges for the \textit{stuck motion}. 

For better comparison, we rescale the values for $t_\text{ret}$ by the constriction length $L_n$ and multiply by $0.96L_0$ 
(approximately the length of a cell body), \textit{i.e.}, $t^n_\text{ret} = 0.96L_0t_\text{ret}/L_n$. The replotted data is presented in 
the inset of Fig.~\ref{fig:ret}. Here, we observe that the $t^n_\text{ret}$ curves for constriction lengths of $0.96L_0$ and $1.44L_0$ collapse 
onto each other. This is understandable, because when the constriction length is comparable to the cell body length, the \tb\ must adapt 
its entire shape to fit into the constriction in order to pass through. We will see below that for these lengths, we also do not observe the 
\textit{stuck-slip} motion. In contrast, shorter constriction lengths with $L_n < L_0$ result in a lower travel speed through the constriction, as 
indicated by higher values of $t^n_\text{ret}$. This may be due to the fact that the rear part (the posterior end) of the parasite still protrudes 
out of the constriction when the front part (the anterior end) already begins to exit. Thus, we speculate that the speed-up mechanism 
discussed earlier when exiting the constriction is counteracted by this slow-down mechanism. The situation becomes relevant when 
the \tb\ needs to passes through constrictions shorter than their body length.
%
%

\begin{figure}
\centering
         \includegraphics[width=0.55\columnwidth,trim={0 0 0 0},clip]{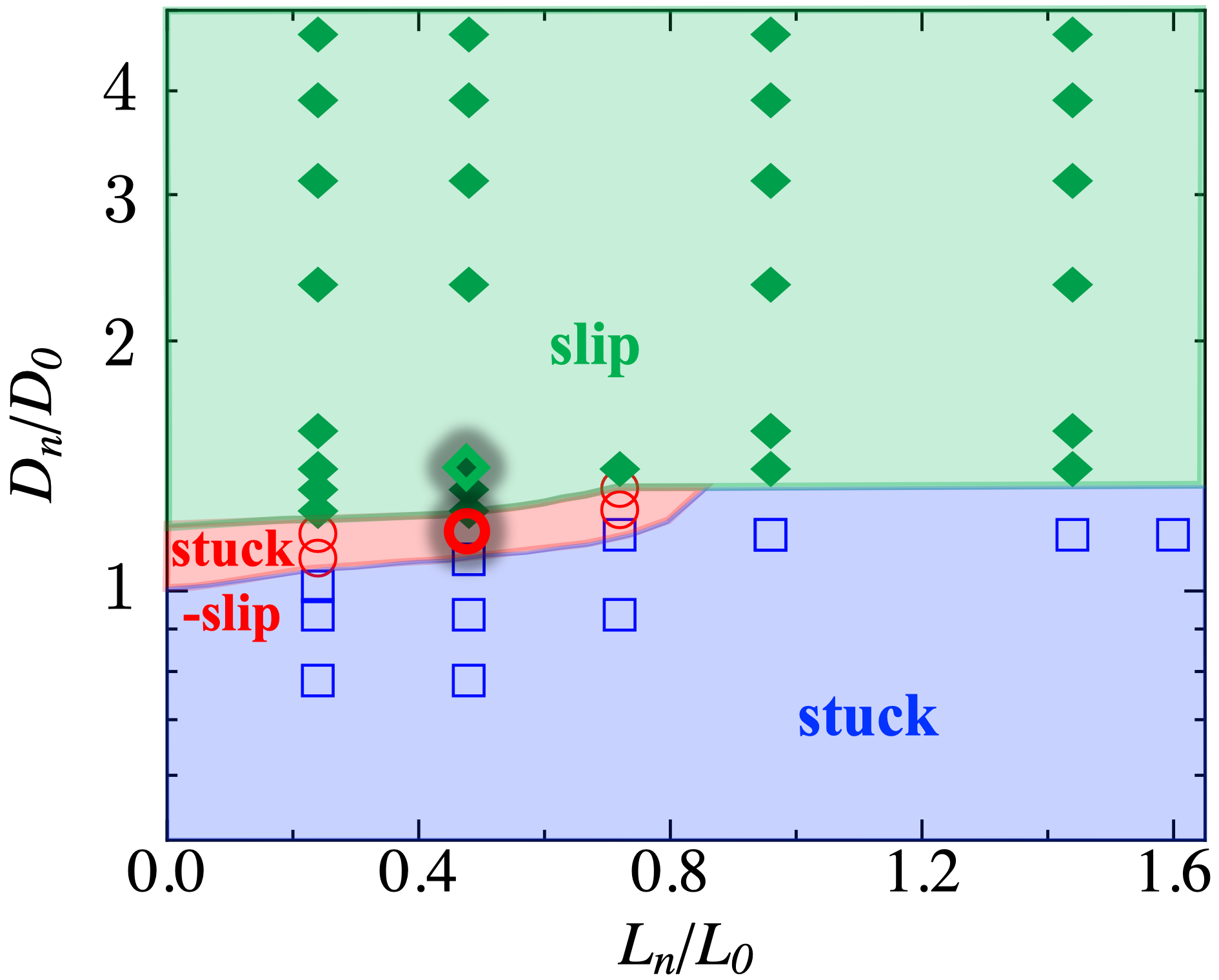}
\caption{\label{fig:states} State diagram of a swimming simulated
\tb\ in microchannels with constrictions. The 
$44$ parameter combinations of $(L_n, D_n)$ are marked according to the identified motional
states: \textit{slip} (green filled diamonds), \textit{stuck-slip} (red open circles), and \textit{stuck} (blue open squares). The two highlighted state 
points correspond to the examples of \textit{slip} and \textit{stuck-slip} motions discussed above and presented in Figs.\ \ref{fig:det_ret} 
and \ref{fig:stuck-slip}, respectively. To emphasize the \textit{stuck-slip} region, the $D_n/D_0$ values are plotted on a logarithmic scale.
}
\end{figure}

\paragraph{State diagram}
We summarize our exploration of the $(L_n,D_n)$ parameter space for the \tb\ swimming in microchannels with constrictions by 
introducing a state diagram that comprises $44$ state points. To draw the state diagram in Fig.~\ref{fig:states}, we work with the following definitions. The \textit{stuck-slip motion} as discussed in 
Fig.\ \ref{fig:stuck-slip} clearly shows time intervalls, where $v_x$ drops to zero and the trajectory exhibits a plateau in $x(t)$. This \textit{stuck-slip} state is
represented as red open circles in the state diagram. The \textit{slip} state, shown as green filled diamonds and discussed in Sec.~\ref{sec:slip},
represents the (almost) smooth passage of the \tb\ through the constriction. This is evident from the trajectories of the anterior and posterior 
ends as well as the CM. In particular, the instantaneous velocity $v_x$ never fully drops to zero.
Finally, the \textit{stuck} state, shown as blue open squares, occurs when the forward motion of the \tb\ is completely hindered and its cell body becomes stuck inside the constriction for the entire simulation time after entering the constriction.

Overall, the \textit{slip} and \textit{stuck} states are relatively intuitive. The \tb\ is expected to pass through constrictions with a width $D_n$ wider than approximately $D_0$ (the widest lateral part of the cell body).  As $D_n$ narrows, the passage becomes blocked. We see this direct transition at roughly $L_n > L_0$ and at a width slightly above $D_0$. It occurs because the straightened cell body immersed inside the constriction, including the beating thinner anterior end, cannot generate sufficient forward hydrodynamic thrust. Interestingly, at $L_n$ below $L_0$ the intermediate \textit{stuck-slip} state appears and the \textit{stuck-slip} region broadenes towards smaller $D_n$ for decreasing $L_n$. As explained before, the anterior end of the cell can protrude from the constriction and through interactions with the wall of the cone region generate a thrust to pull the cell body out of the constriction. This mechanism is more efficient for smaller $L_n$.

\section{Discussions and conclusions}
\label{sec:disc+conc}

Based on a model for an \textit{in silico} \tb\ proposed in our earlier works, we have investigated the locomotion of \tb\ using MPCD-MD simulations in three types of fluid environments: bulk fluid, straight cylindrical microchannels, and microchannels with constrictions. In particular, we have emphasized the distinct dynamics of \tb\ observed in channels with constrictions.

For \tb\ swimming in a bulk viscous fluid, we have conducted a detailed analysis of its swimming trajectory, which exhibits a curved helical path for various sets of mechanical parameters. The mean-squared displacement (MSD) of the center of mass of the \tb\ demonstrates a diffusive-to-ballistic transition characteristic of classical active systems 
for times smaller than the orientational correlation time. However, it also features an intermediate superdiffusive-to-subdiffusive crossover due to the helical nature of the swimming trajectory. In particular, the MSD relative to the local centers of the helical trajectory exhibits damped oscillations. Our analysis indicates that the averaged helical diameter $D_\text{helix}$ is independent of the angular frequency of the flagellar bending wave.
Albeit a specific rule regarding how $D_\text{helix}$ is influenced by the chosen mechanical parameters cannot be established, the results with our standard parameters indicate that the average helix diameter of the center of mass is smaller than that of both the anterior and posterior ends. This finding suggests that the cell body performs a ``kayaking-like" motion while moving forward with a chiral ``crescent-like" shape. Furthermore, our additional analysis of the ``log-rolling" motion reveals that the ratio of flagellar beats per rolling cycle -- an essential parameter for evaluating the accuracy of the \tb\ model -- is comparable to experimental values~\cite{Heddergott:2012,Alizadehrad:2015}, supporting the validity of our parameter selection. 

Furthermore, we have found that the helical swimming trajectory of the \textit{in silico} \tb\ becomes rectified in straight cylindrical channels
compared to the bulk fluid. This effect is first reflected in the constant oscillatory MSD for the relative motion. In particular, the amplitude of the oscillations remains constant over time but decreases as the channel narrows. Geometrical confinement plays a significant role when the 
cylindrical-channel width is comparable to or narrower than the cell-body length. The swimming speed for different channel widths is 
closely linked to the diameter $D_\text{helix}$ of the trajectory. The speed initially increases slightly as the channel narrows and then 
decreases as the helical diameter is compressed. An optimal swimming speed is observed when the channel width is approximately 
twice the bulk helical diameter of the anterior end. It results from an interplay between hydrodynamic interactions, cylindrical confinement, and the high deformability of the parasite. Conversely, the rolling motion inside the channel is enhanced down to a channel 
width of approximately three times the cell's maximum diameter $D_0$. As the channel narrows further, direct steric interactions with 
the walls become dominant, causing the rolling motion to drop to smaller values.

Most importantly, we have explored how the \tb\ swims through microchannels with constrictions. The latter are connected to the 
wider-channel regions by cone-shaped segments. By varying constriction length and width, we could obtain an overall view on the 
behavior of the \tb\ using the  retention time in the constriction and a state diagram that classifies three distinct types of swimming 
motions. Specifically, the \tb\ exhibits a \textit{slip motion} in wider constrictions and a \textit{stuck motion} when the channel width 
reaches the maximum cell-body diameter. Interestingly, when the constriction length becomes shorter than the cell body a peculiar 
\textit{stuck-slip motion} emerges. In this state, the velocity drops to zero and the \tb\ comes to a complete stop so that it spends a significantly longer time inside the constriction. Furthermore, the swimming and shape of the \tb\ as it approaches, navigates through, and exits the constriction is heavily influenced by the dimensions of the constriction and the cone-shaped transition zone. Notably, before entering the constriction, the anterior tip of the \tb\ pushes against the cone-shaped wall and thereby slows down its forward motion. Upon exiting, the anterior tip protudes from the constriction and establishes a pivot on the tilted wall to ``pull" the cell body out of the constriction, which results in an accelerated escape from the constriction. However, the \textit{stuck-slip motion} does not occur, when the constriction length exceeds the cell-body length, as the anterior end cannot effectively protrude beyond the constriction.

The reported simulation study on the \tb\ swimming in various fluid environments provides qualitative insights into how a real \tb\ moves in 
complex biological settings, such as within blood vessels and tissues. The passage through constrictions may be regarded as an initial 
step toward understanding how the \tb\ traverses the blood-brain barrier\ \cite{Mogk:2014,Abbott:2010,Campisi:2018,Park:2019,Hajal:2021}
and it highlights the role the thin anterior end might have in passing such constrictions. Although the model is purely mechanical and 
hydrodynamic, it effectively illustrates the complexity of \tb\ motion through confining geometries. The simulations reveal that a 
mechanical  ``adaptation" emerges in the \tb\ when encountering different environmental conditions. Moreover, our findings also
highlight the role of the cell-body length, when \tb\ swims through both straight cylindrical microchannels and microchannels with constrictions. Specifically, in straight channels it determines the width for which the swimming velocity becomes optimal. More importantly, the \textit{stuck-slip motion} in constrictions arises when the constriction length is shorter than the cell length.

To extend this study and make it more relevant to \tb\ moving in blood flow~\cite{Sun:2018}, a natural progression would be to investigate how the dynamics changes in microchannels with imposed Hagen-Poiseuille flow~\cite{Uppaluri:2012,Zoettl:2012,Zoettl:2013}. Furthermore, as biological tight channels and constrictions, such as blood vessels and interstitial spaces in tissues, are typically deformable rather than rigid, an alternative model should be developed. For example, based on rows and lattices of obstacles, as proposed in~\cite{Heddergott:2012,Muench:2016,Sun:2018}, one could allow for elastic deformations of the obstacles, which then provides a more elaborate representation of the interactions between the \tb\ and its environment.
\section*{Data availability statement}
The data supporting the findings of this study are available from the authors upon reasonable request.
\section*{Author contribution}
HS conceived the research project. ZT and JIUP conducted the numerical simulations and analyzed the data. ZT and HS interpreted the data, participated in the discussions and writing of the manuscript.
\section*{Acknowledgments}
We thank Markus Engstler, Timothy Kr\"uger, Sebastian Rode, Matthias Weiss, and Arne Zantop for helpful discussions. We also gratefully acknowledge the financial support from the DFG project (504947458) and the priority program SPP 2332 “Physics of Parasitism”, which also provided a scientifically stimulating environment.
\section*{Appendix: ``Log-rolling" angle of the cell body}
\label{sec:apps}


\begin{figure} 
	\centering
		\includegraphics[width=0.3\columnwidth,trim={0 0 0 0},clip]{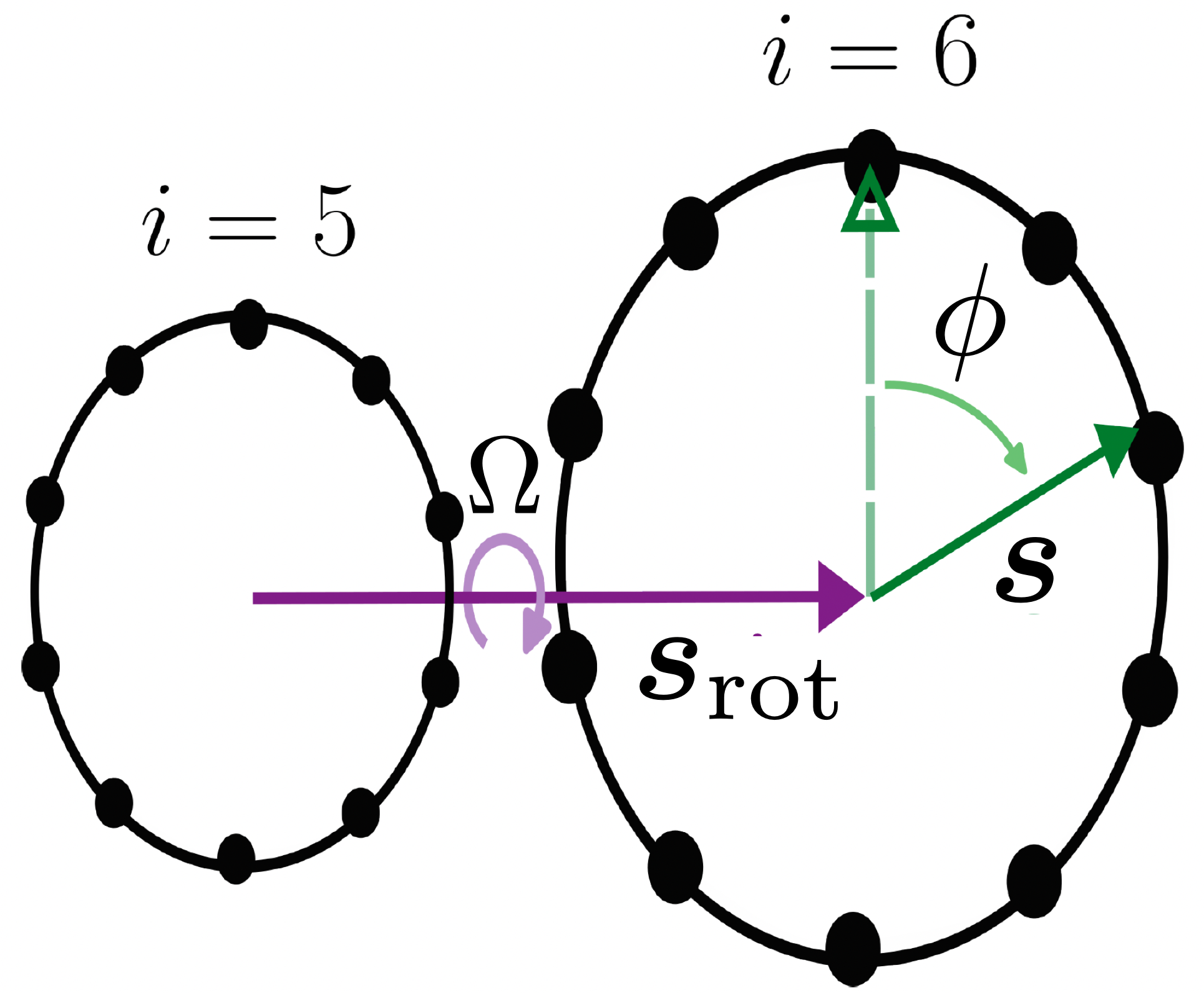}
	\caption{\label{fig:rolling}A graphical depiction of how the body rolling frequency $\Omega$ (light purple) is determined. On 
		the 6th circular segment, the vector $\boldsymbol{s}$ (dark green), connecting the center of the circle to a specific vertex, 
		rotates about the vector $\boldsymbol{s}_\text{rot}$ (dark purple), connecting the centers of both circles, and thereby swipes 
		an angle $\phi$ (light green). Vectors $\boldsymbol{s}$ and $\boldsymbol{s}_\text{rot}$ are here not depicted to scale. 
		Both are normalized to $1$ for the concrete calculations.
	}
	\end{figure} 
	
	The cell body's ``log-rolling" is quantified by an angle $\phi$, defined by a vector $\boldsymbol{s}$ rotating
	around axis $\boldsymbol{s}_\text{rot}$ (see Fig.\ \ref{fig:rolling}). Here, $\boldsymbol{s}$ represents the normalized radius vector at the body's thickest (the $6^\text{th}$)
	section, with $\boldsymbol{s} =\boldsymbol{r}_6/r_6$, where $r_6 = |\boldsymbol{r}_6| $ (see Eq.~\eqref{eq:R_i}). The unit vector $\boldsymbol{s}_\text{rot}$ corresponds to the vector connecting the centers of segments $5$ and $6$. 
	
	To compute the cumulative angle $\phi$ of cell body rotation, we employ an iterative approach. The difficulty is that the whole \tb\ moves 
	around while rotating about its long axis. We identify a specific vertex on the 6th circular segment with its vector $\boldsymbol{s}$ and 
	determine the new orientation of vector $\boldsymbol{s}$ during the motion of the \tb\ after some
	time interval $\delta t$. The condition $\delta t \ll t_\omega$ ensures that the rotational axis $\boldsymbol{s}_\text{rot}$ does not change.
	This results in a sequence of vectors $\{\boldsymbol{s}_1, \boldsymbol{s}_2, ..., \boldsymbol{s}_N\}$ and their corresponding rotation axes $\{\boldsymbol{s_1^{\text{rot}}}, \boldsymbol{s_2^{\text{rot}}}, ..., \boldsymbol{s_{N-1}^{\text{rot}}}\}$. We set the starting angle: $ \phi_0 = 0$.

For each pair of consecutive vectors, we compute the incremental angle $\Delta\phi_i$ through gradient descent optimization. This involves minimizing the squared difference between the rotated vector $\boldsymbol{s}_i$ and the target vector $\boldsymbol{s}_{i+1}$:
\begin{align} \label{eq:roll1}
\Delta\phi_i = \text{argmin}_{\phi} \| \boldsymbol{\Re}(\phi, \boldsymbol{s_i^{\text{rot}}})\boldsymbol{s}_i - \boldsymbol{s}_{i+1} \|^2,
\end{align}
where $\boldsymbol{\Re}(\phi, \boldsymbol{s}_i^\text{rot})$ is the rotation matrix around axis $\boldsymbol{s}_i^\text{rot}$ by angle $\phi$. 

The optimization is performed iteratively using gradient descent. Starting from an initial guess $\phi^{(0)}$, we update the angle according to:
\begin{align} \label{eq:roll2}
\phi^{(k+1)} = \phi^{(k)} - \eta' \frac{\partial}{\partial \phi} \| \boldsymbol{\Re}(\phi^{(k)}, \boldsymbol{s_i^{\text{rot}}})\boldsymbol{s}_i - \boldsymbol{s}_{i+1} \|^2,
\end{align}
where $\eta'$ is the learning rate and $k$ is the iteration index. This process continues until either a maximum number of iterations is reached or the change in $\phi$ becomes smaller than a specified tolerance. 

\begin{figure}[h]
	\centering
		\includegraphics[height=0.45\textwidth,trim={0 0 0 0},clip]{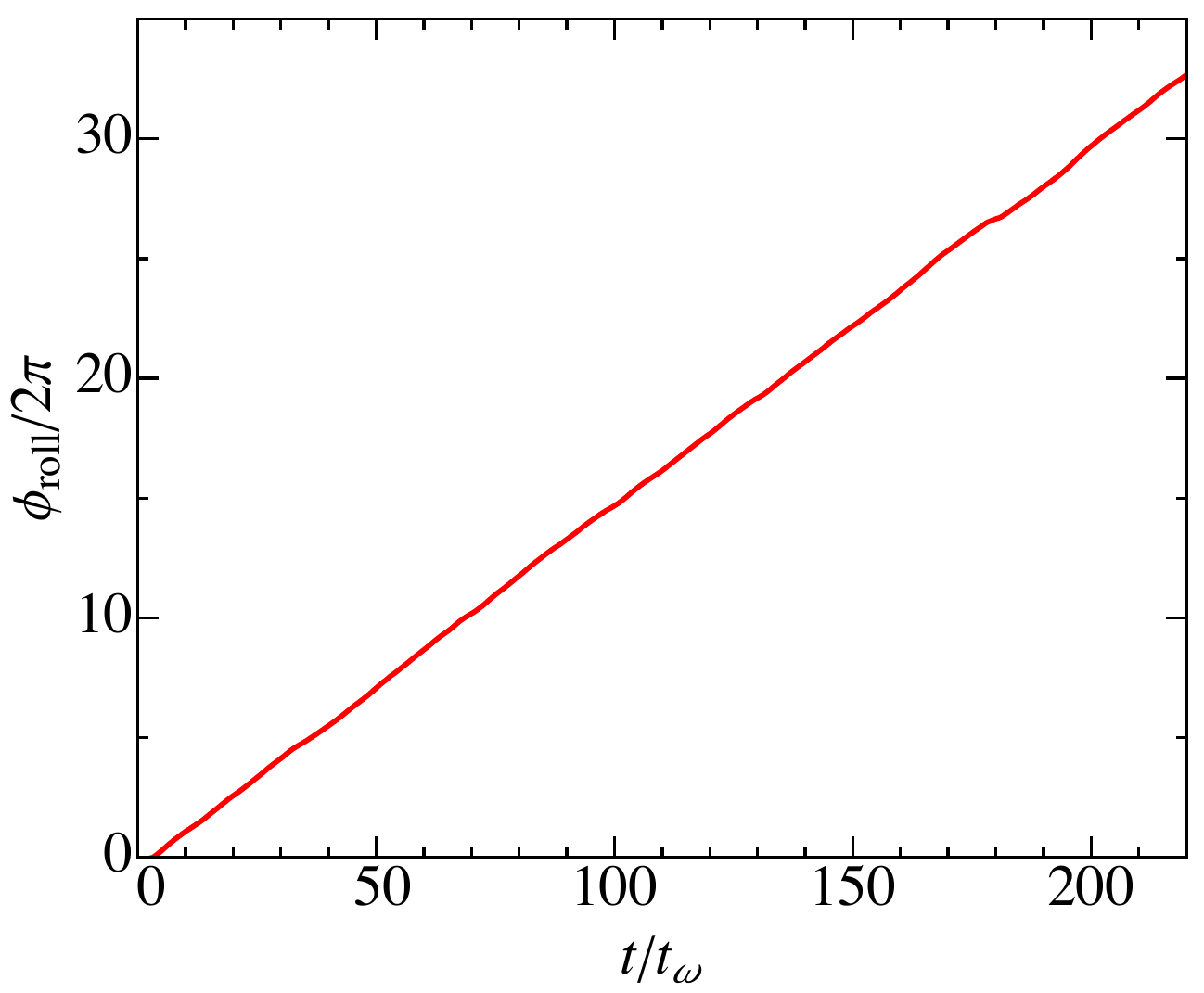}
\caption{\label{fig:Omg}Time evolution of the cell body's rolling angle $\phi$ for the model \tb\ swimming in bulk.
	The standard parameter set of Table~\ref{tb:parameters} is used.
}

\end{figure}

Once the optimal$\Delta\phi_i$ is found, the cumulative angle is updated:
\begin{align} \label{eq:roll3}
\phi_{i}&=\phi_{i-1}+ \Delta\phi_i.
\end{align}
\indent A  typical example of the the rolling angle \textit{vs.} time is shown in Fig.\ \ref{fig:Omg}. A clear linear relation is recognizable. The slope $\text{d}\phi/\text{d}t$ yields the angular frequency $\Omega$ of body rolling.
The calculated average of $\omega/\Omega\approx7$ demonstrates good agreement with experimental measurements of $8\pm2$ flagellar beats per body rotation.

\break
\section*{References}

\begin{thebibliography}{71}
	\expandafter\ifx\csname url\endcsname\relax
	\def\url#1{{\tt #1}}\fi
	\expandafter\ifx\csname urlprefix\endcsname\relax\def\urlprefix{URL }\fi
	\providecommand{\eprint}[2][]{\url{#2}}
	
	\bibitem{Jennings:1901}
	Jennings H~S 1901 {\em Am. Nat.\/} {\bf 35} 369--378
	
	\bibitem{Purcell:1977}
	Purcell E~M 1977 {\em Am. J. Phys.\/} {\bf 45} 3--11
	
	\bibitem{Su:2013}
	Su T~W, Choi I, Feng J, Huang K, McLeod E and Ozcan A 2013 {\em Sci. Rep.\/} {\bf 3} 1664
	
	\bibitem{Jikeli:2015}
	Jikeli J~F, Alvarez L, Friedrich B~M, Wilson L~G, Pascal R, Colin R, Pichlo M, Rennhack A, Brenker C and Kaupp U~B 2015 {\em Nat. Commun.\/} {\bf 6} 7985
	
	\bibitem{Kantsler:2014}
	Kantsler V, Dunkel J, Blayney M and Goldstein R~E 2014 {\em eLife\/} {\bf 3} 02403
	
	\bibitem{Bennett:2015}
	Bennett R~R and Golestanian R 2015 {\em J. R. Soc. Interface\/} {\bf 12} 20141164
	
	\bibitem{Cortese:2021}
	Cortese D and Wan K~Y 2021 {\em Phys. Rev. Lett.\/} {\bf 126} 088003
	
	\bibitem{Leptos:2023}
	Leptos K~C, Chioccioli M, Furlan S, Pesci A~I and Goldstein R~E 2023 {\em Phys. Rev. E\/} {\bf 107} 014404
	
	\bibitem{Marumo:2021}
	Marumo A, Yamagishi M and Yajima J 2021 {\em Commun. Biol.\/} {\bf 4} 1209
	
	\bibitem{Berg:1973}
	Berg H~C and Anderson R~A 1973 {\em Nature\/} {\bf 245} 380--382
	
	\bibitem{Silverman:1974}
	Silverman M and Simon M 1974 {\em Nature\/} {\bf 249} 73--74
	
	\bibitem{Berg:2004}
	Berg H~C 2004 {\em E. coli in Motion\/} (Springer)
	
	\bibitem{Hu:2024}
	Hu J, Gui C, Mao M, Feng P, Liu Y, Gong X and Gompper G 2024 (\textit{Preprint} \eprint{arXiv:2409.13350})
	
	\bibitem{Langousis:2014}
	Langousis G and Hill K~L 2014 {\em Nat. Rev. Microbiol.\/} {\bf 12} 505--518
	
	\bibitem{Rodriguez:2009}
	Rodríguez J~A, Lopez M~A, Thayer M~C, Zhao Y, Oberholzer M, Chang D~D, Kisalu N~K, Penichet M~L, Helguera G, Bruinsma R, Hill K~L and Miao J 2009 {\em Proc. Natl Acad. Sci. USA\/} {\bf 106} 19322--19327
	
	\bibitem{Koyfman:2011}
	Koyfman A~Y, Schmid M~F, Gheiratmand L, Fu C~J, Khant H~A, Huang D, He C~Y and Chiu W 2011 {\em Proc. Natl Acad. Sci. USA\/} {\bf 108} 11105--11108
	
	\bibitem{Heddergott:2012}
	Heddergott N, Krüger T, Babu S~B, Wei A, Stellamanns E, Uppaluri S, 
	Pfohl T, Stark H and Engstler M 2012 {\em PLoS Pathog.\/} {\bf 8} 1--17
	
	\bibitem{Alizadehrad:2015}
	Alizadehrad D, Krüger T, Engstler M and Stark H 2015 {\em PLoS 
		Comput. Biol.\/} {\bf 11} 1--13
	
	\bibitem{Wheeler:2017}
	Wheeler R~J 2017 {\em PLoS Comput. Biol.\/} {\bf 13} 1--22
	
	\bibitem{Shimogawa:2018}
	Shimogawa M~M, Ray S~S, Kisalu N, Zhang Y, Geng Q, Ozcan A and Hill K~L 2018 {\em Sci. Rep.\/} {\bf 8} 9122
	
	\bibitem{Bargul:2016}
	Bargul J~L, Jung J, McOdimba F~A, Omogo C~O, Adung'a V~O, Kr{\"u}ger T, Masiga D~K and Engstler M 2016 {\em PLoS Pathog.\/} {\bf 12} e1005448
	
	\bibitem{Doro:2019}
	D\'or\'o E, Jacobs S~H, Hammond F~R, Schipper H, Pieters R~P, Carrington M,
	Wiegertjes G~F and Forlenza M 2019 {\em eLife\/} {\bf 8} e48388
	
	\bibitem{DeNiz:2023}
	De~Niz M, Frachon E, Gobaa S and Bastin P 2023 {\em PLoS One\/} {\bf 18} 1--31
	
	\bibitem{Sun:2018}
	Sun S~Y, Kaelber J~T, Chen M, Dong X, Nematbakhsh Y, Shi J, Dougherty M, Lim
	C~T, Schmid M~F, Chiu W and He C~Y 2018 {\em Proc. Natl. Acad. Sci. USA\/} {\bf 115} E5916--E5925
	
	\bibitem{Simarro:2012}
	Simarro P~P, Cecchi G, Franco J~R, Paone M, Diarra A, Ruiz-Postigo J~A, Fèvre
	E~M, Mattioli R~C and Jannin J~G 2012 {\em PLoS Negl. Trop. Dis.\/} {\bf 6} 1--12
	
	\bibitem{Mogk:2014}
	Mogk S, Meiwes A, Boßelmann C~M, Wolburg H and Duszenko M 2014 {\em Trends
		Parasitol.\/} {\bf 30} 470--477
	
	\bibitem{Schuster:2017}
	Schuster S, Krüger T, Subota I, Thusek S, Rotureau B, Beilhack A and Engstler
	M 2017 {\em eLife\/} {\bf 6} e27656
	
	\bibitem{Krueger:2018}
	Krüger T, Schuster S and Engstler M 2018 {\em Trends Parasitol.\/} {\bf 34} 1056--1067
	
	\bibitem{Goodwin:1970}
	Goodwin L 1970 {\em Trans. R. Soc. Trop. Med. Hyg.\/} {\bf 64} 797--812
	
	\bibitem{Trindade:2016}
	Trindade S, Rijo-Ferreira F, Carvalho T, Pinto-Neves D, Guegan F, Aresta-Branco
	F, Bento F, Young S, Pinto A, Van Den Abbeele J, Ribeiro R, Dias S, Smith T
	and Figueiredo L 2016 {\em Cell Host Microbe\/} {\bf 19} 837--848
	
	\bibitem{Capewell:2016}
	Capewell P, Cren-Travaillé C, Marchesi F, Johnston P, Clucas C, Benson R~A,
	Gorman T~A, Calvo-Alvarez E, Crouzols A, Jouvion G, Jamonneau V, Weir W,
	Stevenson M~L, O'Neill K, Cooper A, Swar N~r~K, Bucheton B, Ngoyi D~M,
	Garside P, Rotureau B and MacLeod A 2016 {\em eLife\/} {\bf 5} e17716
	
	\bibitem{Reuter:2023}
	Reuter C, Hauf L, Imdahl F, Sen R, Vafadarnejad E, Fey P, Finger T, Jones N~G,
	Walles H, Barquist L, Saliba A~E, Groeber-Becker F and Engstler M 2023 {\em Nat. Commun.\/} {\bf 14} 7660
	
	\bibitem{Abbott:2010}
	Abbott N~J, Patabendige A~A, Dolman D~E, Yusof S~R and Begley D~J 2010 {\em Neurobiol. Dis.\/} {\bf 37} 13--25
	
	\bibitem{Campisi:2018}
	Campisi M, Shin Y, Osaki T, Hajal C, Chiono V and Kamm R~D 2018 {\em Biomaterials\/} {\bf 180} 117--129
	
	\bibitem{Park:2019}
	Park T~E, Mustafaoglu N, Herland A, Hasselkus R, Mannix R, FitzGerald E~A,
	Prantil-Baun R, Watters A, Henry O, Benz M, Sanchez H, McCrea H~J, Goumnerova
	L~C, Song H~W, Palecek S~P, Shusta E and Ingber D~E 2019 {\em Nat. Commun.\/} {\bf 10} 2621
	
	\bibitem{Hajal:2021}
	Hajal C, Le~Roi B, Kamm R~D and Maoz B~M 2021 {\em Annu. Rev. Biomed. Eng.\/} {\bf 23} 359--384
	
	\bibitem{Mogk:2017}
	Mogk S, Boßelmann C~M, Mudogo C~N, Stein J, Wolburg H and Duszenko M 2017 {\em Biol. Rev.\/} {\bf 92} 1675--1687
	
	
	\bibitem{Babu:2012}
	Babu S~B and Stark H 2012 {\em New J. Phys.\/} {\bf 14} 085012
	
	\bibitem{MPCD}
	Gompper G, Ihle T, Kroll D and Winkler R 2009 {\em Adv. Polym. Sci.\/} {\bf 221} 1--87
	
	\bibitem{Overberg:2024}
	Overberg F~A, Jamshidi~Khameneh N, Kr{\"u}ger T, Engstler M, Gompper G and
	Fedosov D~A 2024 (\textit{Preprint} \eprint{bioRxiv:2024.09.27.615450})
	
	\bibitem{Hemphill:1991}
	Hemphill A, Lawson D and Seebeck T 1991 {\em J.\ Parasitol.\/} {\bf 77} 603
	
	\bibitem{Yang:2008}
	Yang Y, Elgeti J and Gompper G 2008 {\em Phys. Rev. E\/} {\bf 78}(6) 061903
	
	\bibitem{kap99}
	Malevanets A and Kapral R 1999 {\em J. Chem. Phys.\/} {\bf 110} 8605
	
	\bibitem{kap00}
	Malevanets A and Kapral R 2000 {\em J. Chem. Phys.\/} {\bf 112} 7260--7269
	
	\bibitem{kapral_review}
	Kapral R 2008 {\em Adv. Chem. Phys\/} {\bf 140} 89--146
	\bibitem{Ripoll:2005}
		Ripoll M, Mussawisade K, Winkler R~G and Gompper G 2005 {\em Phys. Rev. E\/} {\bf 72} 016701
\bibitem{Padding:2004} Padding J T and Louis A A 2004 {\em Phys. Rev. Lett.\/} {\bf 93} 220601
\bibitem{Ripoll:2008}
		Ripoll M, Holmqvist P, Winkler R G, Gompper G, Dhont J K G and Lettinga M P 2008 {\em Phys. Rev. Lett.\/} {\bf 101} 168302
		\bibitem{Franosch:2011}
		Franosch T, Grimm M, Belushkin M, Mor F M, Foffi G, Forr{\'o} L and Jeney S 2011{\em Nature\/} {\bf 478} 85-88 
		\bibitem{Yang:2016}
			Yang M and Ripoll M 2016 {\em Soft Matter\/} {\bf 12} 8504-853
\bibitem{tan01}
		Tan Z, Yang M and Ripoll M 2017 {\em Soft Matter\/} {\bf 13} 7283-7291

\bibitem{tan03}
	Tan Z, Calandrini V, Dhont J~K~G, Nägele G and Winkler R~G 2021 {\em 
		Soft Matter\/} {\bf 17} 7978--7990
\bibitem{Echeverria:2012}
		Echeverria C and Kapral M 2012 {\em Phys. Chem. Chem. Phys.\/} {\bf 14} 6755-6763
\bibitem{Bucciarelli:2016}
		Bucciarelli S, Myung J~S, Farago B, Das S, Vliegenthart G. A., Holderer O, Winkler R G, Schurtenberger P, Gompper G and Stradner A 2016 {\em Sci. Adv.\/} {\bf 2} e1601432
\bibitem{McWhirter:2009}
		McWhirter J~L, Noguchi H and Gompper G 2009 {\em Proc. Natl Acad. Sci. USA\/} {\bf 106} 6039--6043
\bibitem{Dasanna:2019}
		Dasanna A~K, Fedosov D~A, Gompper G and Schwarz U~S 2019 {\em Soft Matter\/} {\bf 15} 5511-5520
\bibitem{Noguchi:2005}
		Noguchi H and Gompper G 2005 {\em Phys. Rev. Lett.\/} {\bf 93} 258102
\bibitem{Weiss:2019} Weiss L B, Marenda M, Micheletti C and Likos C N 2019 {\em 
			Macromolecules\/} {\bf 52} 4111-4119
\bibitem{Huang:2010}
        Huang C C, Winkler R G, Sutmann G and Gompper G 2010 {\em 
	Macromolecules\/} {\bf 43} 10107-10116
\bibitem{Lamura:2019}
	Lamura A and Winkler R G2022 {\em Polymers\/} {\bf 11} 737

\bibitem{Chelakkot:2012}
	Chelakkot R, Winkler R~G and Gompper G 2012 {\em Phys. Rev. Lett.\/} {\bf 109} 178101
\bibitem{ClopesLlahi:2022}
	Clop\'es Llah\'i J, Mart\'in-G\'omez A, Gompper G	and Winkler R G	2022 {\em 
		Phys. Rev. E\/} {\bf 105} 015310
\bibitem{Jaiswal:2024}
		Jaiswal S, Ripoll M and Thakur S 2024 {\em Phys. Chem. Chem. Phys.\/} {\bf 57} 6968-6978

\bibitem{tan02}
	Tan Z, Yang M and Ripoll M 2019 {\em Phys. Rev. Applied\/} {\bf 11} 054004

	
\bibitem{zottl2014}
	Z\"ottl A and Stark H 2014 {\em Phys. Rev. Lett.\/} {\bf 112} 118101
\bibitem{Hu:2015}
	Hu J, Wysocki, M, Winkler R G and Gompper G 2015 {\em Sci. Rep.\/} {\bf 5} 9586
\bibitem{Blaschke:2016}
	Blaschke J, Maurer M, Menon K, Zöttl A and Stark H 2016 {\em Soft Matter\/} {\bf 12} 9821--9831
	
\bibitem{Rode:2019}
	Rode S, Elgeti J and Gompper G 2019 {\em New J. Phys.\/} {\bf 21} 013016
\bibitem{Rode:2021}
		Rode S, Elgeti J and Gompper G 2021 {\em Eur. Phys. J. E\/} {\bf 44} 76

\bibitem{Zantop:2022}
	Zantop A~W and Stark H 2022 {\em Soft Matter\/} {\bf 18} 6179--6191
\bibitem{Roca:2022}
		Roca-Bonet S, Wagner M and Ripoll M 2022 {\em Soft Matter\/} {\bf 18} 7741-7751
\bibitem{Clopes:2022}
		Clop\'es J, Winkler R G and Gompper G 2022 {\em J. Chem. Phys.\/} {\bf 156} 194901
\bibitem{Ruehle:2020} R\"uhle F and Stark H 2020 {\em Eur. Phys. J. E\/} {\bf 43} 26
\bibitem{Ning:2023}
	Ning L, Lou X, Ma Q, Yang Y, Luo N, Chen K, Meng F, Zhou X, Yang M and Peng Y 2023 {\em Phys. Rev. Lett.\/} {\bf 131} 158301
\bibitem{McGovern:2024}McGovern A D, Huang M-J, Wang J, Kapral R and Aranson I S  2024 {\em Small\/} {\bf 20} 2304773
	
	\bibitem{Allahyarov:2002}
	Allahyarov E and Gompper G 2002 {\em Phys. Rev. E\/} {\bf 66} 036702
	
	\bibitem{ihl03}
	Ihle T and Kroll D~M 2003 {\em Phys. Rev. E\/} {\bf 67} 066706
	
	\bibitem{allen}
	Allen M~P and Tildesley D~J 1989 {\em Computer Simulation of Liquids\/} (Oxford
	University Press)
		
	\bibitem{padding2005stick}
	Padding J, Wysocki A, L{\"o}wen H and Louis A 2005 {\em J. Phys.: Condens. Matter\/} {\bf 17} S3393
	
	\bibitem{padding06}
	Padding J~T and Louis A~A 2006 {\em Phys. Rev. E\/} {\bf 74} 031402
	
	\bibitem{OlartePlata:2019}
	Olarte-Plata J~D and Bresme F 2019 {\em Phys. Chem. Chem. Phys.\/} {\bf 21} 1131--1140
	
	\bibitem{Muench:2016}
	Münch J~L, Alizadehrad D, Babu S~B and Stark H 2016 {\em Soft Matter\/} {\bf 12} 7350--7363
	
	\bibitem{theers2016modeling}
	Theers M, Westphal E, Gompper G and Winkler R~G 2016 {\em Soft Matter\/} {\bf 12} 7372--7385
		
	\bibitem{Kai:2022}
	Qi K, Westphal E, Gompper G and Winkler R~G 2022 {\em Commun. Phys.\/} {\bf 5} 49
	
	\bibitem{hecht2005}
	Hecht M, Harting J, Ihle T and Herrmann H~J 2005 {\em Phys. Rev. E\/} {\bf 72} 011408
	
	\bibitem{Zoettl:2012}
	Z\"ottl A and Stark H 2012 {\em Phys. Rev. Lett.\/} {\bf 108} 218104
	
	\bibitem{Schaar:2015}
	Schaar K, Z\"ottl A and Stark H 2015 {\em Phys. Rev. Lett.\/} {\bf 115} 038101
	
	\bibitem{Li:2009}
	Li S and Jain A 2009 {\em Encyclopedia of Biometrics: I - Z.\/}  (Springer) 
\bibitem{Pays23} Pays E, Radwanska M and Magez S 2023 {\em Annu. Rev. Pathol. Mech. Dis.\/} {\bf18} 19-45

	\bibitem{Uppaluri:2012}
	Uppaluri S, Heddergott N, Stellamanns E, Herminghaus S, Z\"ottl A, Stark H, Engstler M and Pfohl T 2012 {\em Biophys. J.\/} {\bf 103} 1162--1169
	
	\bibitem{Zoettl:2013}
	Z{\"o}ttl A and Stark H 2013 {\em Eur. Phys. J. E\/} {\bf 36} 4
	\bibitem{Vizsnyiczai:2020}
		Vizsnyiczai G, Frangipane G, Bianchi S, Saglimbeni F, Dell’Arciprete D and Di Leonardo R 2020 {\em Nat. Commun.\/} {\bf 11} 2340
\end{thebibliography}
\providecommand{\newblock}{}

\end{document}